\documentclass[12pt]{article}
\usepackage[top=1.25in,bottom=1.25in,left=1.25in,right=1.25in]{geometry}
\usepackage[utf8]{inputenc}
\usepackage{array,multirow}
\usepackage[toc,page,header]{appendix}
\usepackage{subfig}
\usepackage{booktabs}
\usepackage{setspace}
\usepackage{amsmath}
\usepackage{amssymb}
\usepackage{amsthm}
\usepackage{xcolor}
\usepackage{graphicx}
\usepackage{float}
\usepackage{enumerate}
\usepackage[utf8]{inputenc}
\usepackage{bbm}
\usepackage{bm}
\usepackage{appendix}

\usepackage{mathrsfs}
\usepackage{comment}
\usepackage{apptools}
\usepackage[linkcolor=blue]{hyperref}

\usepackage{natbib}
\usepackage{setspace,lipsum}

\newcommand{\indep}{\perp \!\!\! \perp}
\providecommand{\eps}{\varepsilon}

\newtheorem{definition}{Definition}
\newtheorem{proposition}{Proposition}
\newtheorem{theorem}{Theorem}

\newtheorem*{corollary*}{Corollary}
\newtheorem{lemma}{Lemma}

\newtheorem{assumption}{Assumption}

\theoremstyle{remark}
\newtheorem{example}{Example}
\theoremstyle{remark}

\AtAppendix{
\numberwithin{corollary}{section}
\numberwithin{assumption}{section}
\numberwithin{equation}{section}
\numberwithin{definition}{section}
\numberwithin{theorem}{section}
\numberwithin{proposition}{section}
\numberwithin{lemma}{section}}

\title{Data and Incentives\thanks{We are grateful to Eduardo Azevedo, Cuimin Ba, Dirk Bergemann, Alessandro Bonatti, Sylvain Chassang, Yash Deshpande, Ben Golub, Matt Jackson, Yizhou Jin, Navin Kartik, Rishabh Kirpalani, Alessandro Lizzeri, Steven Matthews, Xiaosheng Mu, Larry Samuelson, Andrzej Skrzypacz, Juuso Toikka, and Weijie Zhong for useful conversations, and to National Science Foundation Grant SES-1851629 for financial support.  We thank Changhwa Lee for valuable research assistance on this project.}}
\author{Annie Liang\footnote{Department of Economics and Department of Computer Science, Northwestern University} \quad \quad \quad Erik Madsen\footnote{Department of Economics, New York University}}

\date{September 1, 2022}

\begin{document}
\maketitle

\begin{abstract}

``Big data" gives markets access to previously unmeasured characteristics of individual agents. Policymakers must decide whether and how to regulate the use of this data. We study how new data affects incentives for agents to exert effort in settings such as the labor market, where an agent's quality is initially unknown but is forecast from an observable outcome. We show that measurement of a new covariate has a systematic effect on the average effort exerted by agents, with the direction of the effect determined by whether the covariate is informative about long-run quality or about a shock to short-run outcomes. For a class of covariates satisfying a statistical property we call \emph{strong homoskedasticity}, this effect is uniform across agents. More generally, new measurements can impact agents unequally, and we show that these distributional effects have a first-order impact on social welfare. 

\end{abstract}
\bigskip

\section{Introduction}

Online platforms and data brokers extensively track, record, and aggregate consumer activities, producing measurements of everything from the size of an individual's social network,\footnote{The finance startup Lenddo evaluated borrowers on the basis of factors such as ``how many friends or followers they have on their social networks'' (\url{https://www.wsj.com/articles/BL-DGB-24461}).} to how often they move residences,\footnote{The alternative credit scoring company ZestFinance used borrowers' frequency of residence changes to predict their creditworthiness (\url{https://tinyurl.com/3ay8zhw7}).} to the amount of time they spend playing video games.\footnote{China's widely-publicized social credit scoring system reportedly plans to incorporate data on how many video games a consumer purchases and how much time they spend playing them (\url{ tinyurl.com/m55aapz8 }).} These new measurements are increasingly  available to firms and organizations, who may find them useful as predictors of economic outcomes---e.g., of a worker's future productivity in a new job.\footnote{Employers already widely use similar information collected from internet searches to screen potential hires on the basis of factors such as social media activity (\url{https://tinyurl.com/p22p2pch}).}  Regulating such uses of personal data has emerged as an important policy issue,\footnote{For instance, proposed European Union rules for artificial intelligence have flagged automated employment screening systems as ``high risk'' applications subject to strict regulation, in particular regarding the datasets they rely on (\url{https://ec.europa.eu/commission/presscorner/detail/en/ip_21_1682}).} but our understanding of when and how to do so remains preliminary.  The use of new data in such settings may have far-ranging social impacts beyond direct privacy concerns, reshaping the creation and distribution of economic surplus.

In this paper, we study the impact of new data on markets in which moral hazard is an important concern.\footnote{Our approach complements recent work focusing on how data collection impacts markets shaped by asymmetric information. See, for instance, \citet{BergemannBonattiGan, ElliottGaleotti, YangJMP} for price discrimination; \citet{Ichihashi,HidirVellodi,GomesPavan} for matching on platforms; and \citet{Chassang,LambaSegura} for insurance pricing.}  A motivating application is the labor market, where wages and job opportunities are commonly tied to market forecasts of a worker's productivity based on past output. A well-recognized consequence of this practice is that workers are incentivized to work hard to improve the market's forecast.
Since output is typically socially valuable, new data which impacts workers' effort incentives may negatively affect labor market productivity absent regulation. 

We propose a simple model of reputational incentives that isolates the effect of new data on moral hazard.  Our model builds on the classic ``career concerns'' framework of \citet{Holmstrom}, in which an agent exerts effort to improve an outcome used by a market to forecast his type. Different from \citet{Holmstrom}, we suppose that the market additionally bases its forecast on auxiliary data consisting of covariates describing the agent, which are observed prior to his choice of effort.  

We separate covariates into two categories: Some covariates, which we call \emph{attributes}, describe the agent's type; while others, which we call \emph{circumstances}, are informative about a transient shock to his outcome.  For example, a worker's creativity is an attribute, while  an illness or injury is a circumstance.  We model the acquisition of new data as an expansion of the set of covariates that are measured and may be used for forecasting. Measurement of new covariates updates the market's beliefs about a given agent's type and shock, reshaping incentives for effort. 

Our main results characterize how measurement of a new covariate impacts the population distribution of effort and aggregate welfare.  Our basic positive result is that incorporating a new covariate into the market's type forecast leads to both a systematic \emph{reduction in uncertainty} across the population as well as a \emph{redistribution of uncertainty} between agents.  While the systematic effect moves the effort of all agents in the same direction, the redistributionary effect leads to heterogeneous effort responses which may differ even directionally across agents.  Each of these effects has a first-order effect on aggregate welfare, and we find that the redistributionary effect can oppose and even overturn the welfare impact of reducing uncertainty. 

We formalize our positive findings through a pair of theorems.  Theorem \ref{thm:gencorr} shows that quite generally, measurement of a new attribute reduces average effort in the population, while measurement of a new circumstance increases average effort.  Importantly, this result does not guarantee that all agents change their effort by the same amount, or even in the same direction.  Theorem \ref{thm:mean-shift} shows that any heterogeneity in the effort responses of different agents is entirely attributable to a redistribution of uncertainty across the population.  
 
Our normative findings are summarized in Theorem \ref{thm:welfare}.  It establishes that, in the absence of redistribution of uncertainty, the directional effect on welfare of a newly measured covariate is jointly determined by its classification as an attribute or circumstance along with the weight that agents place on their future reputations.  It further shows that greater redistribution of uncertainty leads to reduced welfare gains from measurement of a covariate, and that this reduction can be so extreme that measurement of some covariates is \emph{never} welfare-enhancing, regardless of the magnitude of agents' reputational concerns.    

Our work contributes to an emerging literature studying the use of personal data for forecasting. Existing work has highlighted incentives for agents to game forecasts by distorting \citep{Ball, BonattiCisternas, FrankelKartik, Haghtalabetal, Immorlica} or misreporting \citep{EliazSpiegler1,EliazSpiegler2} their covariates.  Such incentives are especially important when a small number of covariates shape forecasts in a well-understood way. We study the complementary question of how data usage impacts incentives for agents to directly improve outcomes.  This incentive is particularly relevant when outcomes are a primary forecasting input, making them a natural target for manipulation; or when algorithms used to incorporate additional covariates into forecasts are opaque, obscuring effective strategies for gaming them.

Methodologically, our model builds on the career concerns literature. Compared to the original \citet{Holmstrom} model, we focus on a two-period model which incorporates auxiliary signals and non-Gaussian information structures.  We share these modeling features with the closely related work of \citet{DJT} (hereafter DJT) and \citet{rodina2018}.\footnote{An adjacent literature on relative performance comparisons, e.g.\ \citet{MeyerVickers}, considers settings in which the additional signal is not exogenous, but is instead generated by the outcome of another agent with correlated unobservables.  See also \citet{Tirole}, in which the additional signal is an outcome in another domain in which the agent exerts effort.}  They focus on an environment in which the market receives signals about a single agent, while our work instead allows for heterogeneous effort across a population of agents.\footnote{Specifically, in our model, agents are differentiated by their covariate values (i.e., signal realizations) which are realized prior to their effort choices.}  This feature yields novel predictions about how new data shapes the population distribution of effort.  The effort dispersion generated by new data has a first-order effect on welfare, and it can even dictate whether a given dataset is welfare-improving or not (see Section \ref{subsec:welfare}).  Our results therefore highlight the importance of explicitly modeling the heterogeneity across agents that is present in applications.

The remainder of this paper proceeds as follows.  Section \ref{sec:Model} describes our model; Section \ref{sec:results} establishes our main results about the impact of new measurements on effort and welfare; Section \ref{sec:extensions} discusses extensions; and Section \ref{sec:conclusion} concludes.  Supporting analyses and all proofs are collected in the Appendix.

\section{Model}  \label{sec:Model}
In Section \ref{sec:holmstrom} we describe our basic model of reputational incentives for effort, which is a 2-period version of the \citet{Holmstrom} career concerns model with general information structures. In Section \ref{sec:data} we augment the model by introducing auxiliary data.

\subsection{Effort and Welfare} \label{sec:holmstrom}

An agent participates in a market across two periods $t=1,2$. He possesses a quality type $\theta \sim F_\theta$, which is persistent across time and unknown to himself and the market.

In \textbf{period 1}, the agent privately chooses an effort level  $e \in \mathbb{R}_+$ at cost $C(e) = \frac12 e^2$. (We extend our results to general cost functions in Section \ref{subsec:genCost}.) The agent's effort choice, along with his quality $\theta$ and a transient shock $\eps \sim F_\eps$, determine the realization of an observable outcome 
\[Y = e + \theta + \eps.\] 
We assume that $\mathbb{E}(\theta) = \mu > 0$ while $\mathbb{E}(\eps) = 0.$  The agent's reward from his period-1 interaction is independent of $Y$ and normalized to be 0.\footnote{This normalization does not rule out wage payments which depend on the market's forecast of period-1 effort, as in \citet{Holmstrom}.  Such effects do not impact equilibrium effort or social surplus, and so we do not explicitly model them.}  His period-1 payoff is therefore:
\[U_1 = - \frac12 e^2.\]

In \textbf{period 2}, the agent receives a reputational payoff standing in for returns from future participation in the market.  This payoff is equal to the market's expectation of his quality conditional on the outcome variable $Y$.\footnote{None of our results would change if the agent's reputational payoff were instead the market's expectation of any strictly increasing function of $\theta$.  Our model therefore accommodates a variety of interpretations for the source of reputational returns from effort.} 
Since the agent's effort choice is private, the market's forecast is based on a conjectured level of effort $\hat{e}$. Letting $Y^{\hat{e}} \equiv \hat{e} + \theta + \eps$ be the outcome supposing that the market's effort conjecture is correct, the agent's second-period payoff conditional on the realized outcome $Y=y$ is
\[
U_2 = \mathbb{E}^{\hat{e}}\left(\theta \mid Y = y \right),\]
where $\mathbb{E}^{\hat{e}}(\theta \mid Y = y)$ denotes the market's (potentially misspecified) expectation of $\theta$, updated based on the realized outcome assuming that $Y = Y^{\hat{e}}$.

The agent's ex-post payoff from participating in the market in both periods is a weighted sum of payoffs across the two periods: \[U = (1 - \beta) \cdot U_1 + \beta \cdot U_2,\] where $\beta \in (0, 1)$ is the \emph{reputation weight}, which denotes the importance to the agent of future reputational rewards versus current effort costs. The agent's expected payoff under effort level $e$ is therefore
\[\mathbb{E}^e(U) = \beta \cdot \mathbb{E}^e(\mathbb{E}^{\hat{e}}(\theta \mid Y)) - (1-\beta) \cdot \frac{e^2}{2},\]
where $\mathbb{E}^e$ denotes the expectation operator given the true effort level $e$. 

In equilibrium, the agent must have no incentive to deviate from the market's conjectured level of effort.  Let $e^*$ denote equilibrium effort.  Then in equilibrium the marginal value of effort (i.e., the equilibrium marginal impact of effort on the expected reputational reward), discounted by its relative weight $\beta/(1 - \beta),$ equals the marginal cost of effort:
\[
\left.\frac{\beta}{1 - \beta} \cdot \frac{\partial}{\partial e} \mathbb{E}^{e}(\mathbb{E}^{e^*}(\theta \mid Y))\right|_{e=e^*} = e^*.\]
Because effort impacts the outcome additively, the marginal value of effort appearing in this first-order condition is independent of $e^*$ and may be written as 
\begin{equation} \label{def:MV}
MV \equiv \mathbb{E}^0\left(\frac{\partial}{\partial Y}\mathbb{E}^0(\theta \mid Y)\right),
\end{equation}
where $\mathbb{E}^0$ denotes the expectation operator assuming that the agent does not exert effort to distort the outcome (see Appendix \ref{app:MVChar} for details).  As (\ref{def:MV}) does not depend on $e^*,$ the unique effort level satisfying the first-order condition is then $
e^* = \frac{\beta}{1 - \beta} \cdot MV.$   Throughout this paper, we will assume that the first-order approach is valid, so that $e^*$ constitutes the unique equilibrium effort choice.

We measure welfare using a standard criterion that treats both the outcome variable $Y$ and the agent's effort cost $C(e)$ as welfare-relevant.\footnote{The assumption that $Y$ contributes to social welfare is appropriate for settings such as the labor market, in which the outcome captures productive output or some other socially valuable activity. We do not include the agent's equilibrium reputational payoff in the welfare calculation because on average that payoff is fixed at $\mathbb{E}^{e^*}(\mathbb{E}^{e^*}(\theta \mid Y)) = \mu$, independent of the equilibrium effort level.  (This property continues to hold when the model is augmented with data.)}  An agent whose type is $\theta$ and effort choice is $e$ thus generates welfare
\begin{align}w(\theta, e) \equiv \mathbb{E}^e(Y \mid \theta) - C(e) = \theta + e - \frac12 e^2.\label{eq:welfareExPostDef}\end{align}
This function  is strictly concave in effort and  maximized at the ``first-best'' effort level $e^{FB} = 1$ no matter the agent's type. Appendix \ref{app:welfareAlt} extends our analysis to alternative welfare specifications incorporating learning-by-doing and unproductive gaming.

\subsection{Data and Beliefs} \label{sec:data} 
We now augment the basic model by supposing that $\theta$ and $\eps$ are predictable from underlying (and potentially measurable) \emph{covariates}, with data revealing a subset of these covariates. We refer to those covariates which predict $\theta$ as \emph{attributes}, denoted by the random variables $(a_1, \dots , a_J)$, and those covariates which predict $\eps$ as \emph{circumstances}, denoted by the random variables $(c_1, \dots, c_K)$. Specifically, the type $\theta$ and shock $\eps$ satisfy
\begin{align*}
    \theta & = f^1(a_1) + \dots + f^J(a_J) + u_\theta \\
    \eps & = g^1(c_1) +  \dots + g^K(c_K) + u_\eps
\end{align*}
where each $f^j$, $j\in \{1,\dots,J\}$, and $g^k$, $k\in \{1,\dots,K\}$, is a deterministic and one-to-one effect size function. (This specification nests the standard linear regression model as a special case when all effect size functions are affine.)  For convenience, we will define the \emph{type components} $\theta_j \equiv f^j(a_j)$ for each $j = 1, ..., J$ and \emph{shock components} $\eps_k \equiv g^k(c_k)$  for each $k = 1, ..., K.$\footnote{Invertibility of the effect size functions implies that observation of a new covariate $a_j$ or $c_k$ is equivalent to observation of the corresponding type or shock component $\theta_j$ or $\eps_k.$ Some of our results, particularly those involving Strong Homoskedasticity, do not depend on invertibility.}

The idiosyncratic noise terms $u_\theta$ and $u_\eps$ are independent of one another and  of all covariates, and have full support on the reals.\footnote{The full support assumption ensures that the distributions of $\theta$ and $\eps$ conditional on any family of measured covariates have full support, simplifying our proofs.  All of our results continue to hold in the absence of full support, and so we will make free use of examples which do not feature full-support idiosyncratic noise terms.} We allow for correlation between attributes and between circumstances, but assume that the vector of attributes is independent of the vector of circumstances, i.e.\ $(a_1, \dots, a_J) \indep (c_1, \dots, c_K)$, implying in particular that $\theta  \indep \eps.$  (We consider covariates which are correlated with both the type and shock in Section \ref{subsec:correlation}.) 

Some covariates are measured, making them observable to the agent and the market. We use $\mathcal{J} \subseteq \{1, \dots, J\}$ to denote the set of measured attributes and $\mathcal{K} \subseteq \{1, \dots, K\}$ to denote the set of measured circumstances.  All measured covariates are observed at the outset of the interaction, leading the agent and market to share a common belief that the agent's type and shock follow their distributions conditional on the agent's measured covariate values.  We view symmetric uncertainty as a natural conceptual benchmark that allows us to cleanly disentangle moral hazard from issues of selection.  In Section \ref{sec:discuss} we discuss how our results would change if either side had additional private information.

The interaction then proceeds as described in Section \ref{sec:holmstrom}, with appropriate adjustments to the calculation of equilibrium effort.   Conditioning on the measured covariates, the agent's marginal value of effort changes from (\ref{def:MV}) to the quantity
\begin{equation} \label{def:MVexpand}
MV_{\mathcal{J},\mathcal{K}} \equiv \mathbb{E}^0\left(\frac{\partial}{\partial Y}\mathbb{E}^0(\theta \mid Y, a_{\mathcal{J}}, c_{\mathcal{K}}) \mid a_{\mathcal{J}}, c_{\mathcal{K}}\right)
\end{equation}
and equilibrium effort becomes
\begin{equation} \label{eq:EffortRandom}
e^*_{\mathcal{J},\mathcal{K}} =  \frac{\beta}{1 - \beta} \cdot MV_{\mathcal{J},\mathcal{K}}.
\end{equation}

Note that both the marginal value of effort and equilibrium effort may vary with the values of the agent's measured covariates, and they are therefore both random quantities.  We interpret this randomness from a population perspective, by supposing that the market interacts with a continuum of agents possessing varying attributes and circumstances.  From this perspective, random variation in $e^*_{\mathcal{J}, \mathcal{K}}$ corresponds to a distribution of effort across the population of agents.

Aggregate welfare given measured covariates $(\mathcal{J},\mathcal{K})$ is the expectation of realized welfare $w\left(\theta, e^*_{\mathcal{J},\mathcal{K}}\right)$ as defined in \eqref{eq:welfareExPostDef}, averaging over variation in the type $\theta$ and effort $e^*_{\mathcal{J},\mathcal{K}}$ across the population:
\[W(\mathcal{J}, \mathcal{K}) \equiv  \mathbb{E}\left( w\left(\theta, e^*_{\mathcal{J},\mathcal{K}}\right) \right) = \mu + \mathbb{E} \left(e^*_{\mathcal{J},\mathcal{K}} - \frac{1}{2} \left(e^*_{\mathcal{J},\mathcal{K}}\right)^2\right).\]
(Recall that $\mu \equiv \mathbb{E}(\theta)$ is the unconditional average quality in the population.) Aggregate welfare is maximized when all agents exert the first-best level $e^{FB} = 1$.

Our main results compare effort and welfare when the set of measured covariates changes from some \emph{baseline} family $(\mathcal{J}, \mathcal{K})$ to an \emph{expanded} family $(\mathcal{J}\cup\{j'\}, \mathcal{K})$ or $(\mathcal{J}, \mathcal{K}\cup\{k'\})$ containing one additional covariate. To simplify exposition, throughout the main text we develop our results assuming that the baseline family is $(\mathcal{J},\mathcal{K}) = (\emptyset,\emptyset)$, while the expanded family is either $(\{1\},\emptyset)$ or $(\emptyset, \{1\})$, corresponding to measurement of attribute 1 or circumstance 1. (Our results extend straightforwardly to general baselines---see Appendix \ref{app:GenBaseline} for details).  In this context, we define $\eta\equiv\sum_{j>1} f^j(a_j) + u_\theta$ and $\delta \equiv \sum_{k>1} g^k(c_k) + u_\eps$ and decompose the type and shock as
\begin{align*}
   \theta =  \theta_1 + \eta, \quad \eps =  \eps_1 + \delta,
\end{align*} 
so that the type component $\theta_1$ and shock component $\eps_1$ summarize the information revealed by a new measurement, while $\eta$ and $\delta$ are the residual unknowns.

We impose a set of standard regularity conditions on the distributions of these variables. 

\begin{assumption}[Admissibility]\label{ass:regularity}
The random variables $\theta$ and $\eps$ have log-concave density functions, and the conditional random variables $\eta \mid a_1$ and $\delta \mid c_1$ have log-concave density functions for every realization of $a_1$ and $c_1$.
\end{assumption}

\begin{assumption}[Differentiability] \label{ass:expectationDiff} For every effort level $e,$ the derivative $\frac{\partial}{\partial Y}\mathbb{E}^e\left(\theta \mid Y\right)$ exists and is uniformly bounded across all realizations of $Y.$  Additionally, for every effort level $e$ and realization of $(a_1, c_1),$ the derivatives
$\frac{\partial}{\partial Y}\mathbb{E}^e\left(\theta \mid Y, a_1\right)$ and $\frac{\partial}{\partial Y}\mathbb{E}^e\left(\theta \mid Y, c_1\right)$ exist and are uniformly bounded across all realizations of $Y$.
\end{assumption}

In additive statistical inference models, log-concavity is a canonical assumption ensuring that better outcomes correspond to improved inferences about latent variables.\footnote{Specifically, if an analyst observes an outcome $Z$ which is decomposable as $Z = X + Y,$ where both $X$ and $Y$ are unobserved, and if $X$ and $Y$ are statistically independent and have log-concave density functions, then upon observing $Z$ his posterior beliefs about $X$ and $Y$ are higher (in the first-order stochastic dominance order) for larger realizations of $Z$ \citep{MilgromMLRP}.} Assumption \ref{ass:regularity} ensures that in both the baseline and the expanded environments, better (worse) realizations of $Y$ lead to higher (lower) posterior beliefs about both the type and shock.
Assumption \ref{ass:expectationDiff} ensures that conditional expectations are sufficiently smooth that we can take derivatives and exchange derivatives and expectations where required.

\subsection{Discussion of Modeling Choices} \label{sec:discuss}

\paragraph{Private information on the side of the market.} Our model assumes that the market does not know more about an agent's type or shock than the agent does himself. In some applications, private information on the side of the market is possible if the market has access to data on past outcomes for other agents with similar covariates.  (This view of big data is embedded in, for instance, the ``inverse selection'' model of \citet{LambaSegura}.) We show in Section \ref{subsec:modelUncertainty} that our results continue to hold under this sort of informational asymmetry, so long as measurement of a new covariate leads the agent to believe that the market has gained new information about his type or shock. 

\paragraph{Private information on the side of the agent.}
Another possibility is that the agent knows more about his type and shock than the market. This asymmetry implies not only that the agent know more about his own covariates, but also that he can discern how these covariates impact his type and shock distributions (via the effect size functions $f^j$ and $g^k$), a demanding assumption in many applications. Nevertheless, our results continue to hold with private information on the agent's side whenever the market's posterior type expectation is linear in the outcome signal---for instance, whenever $(\theta_1, ..., \theta_J)$ and $(\eps_1, ..., \eps_K)$ follow elliptical distributions. Beyond these settings, the agent's perceived marginal value of effort may vary with private information about his type or shock, complicating the analysis, but we conjecture that our main results extend more broadly. 

\paragraph{Exogeneity of covariates.}
Our model contrasts covariates, which are fixed characteristics of the agent (at least in the short run); and the outcome $Y$, which is susceptible to manipulation by effort.  We view this dichotomy as a useful one for several reasons. First, $\theta$ and $\eps$ may be determined by the aggregation of many covariates, each of which individually plays only a small role. In such contexts our exercise can be viewed as focusing on the agent's incentives to influence the relatively informative outcome signal $Y$, while abstracting from any costly distortion of individual less-informative covariates. Second,  to compute the value of manipulating a covariate, the agent must know the precise shape of the effect size function describing how that covariate impacts the outcome, which is more demanding than the knowledge requirements that we impose.

\section{Main Results}\label{sec:results}

Our main results characterize the impact of measuring a new covariate on population effort and welfare.  We first show that for a broad class of covariates, measuring a new attribute decreases population effort on average, while measuring a new circumstance increases it (Section \ref{subsec:affiliated}). However outside a narrower class of covariates, new measurements may yield effort responses of heterogeneous magnitude and even direction across agents (Section \ref{subsec:SH}).  In Section \ref{subsec:welfare} we combine these insights to study the effect of measuring a new covariate on aggregate welfare, taking into account both the resulting average effect on effort as well as the induced heterogeneity.  

\subsection{The Average Impact of New Measurements}\label{subsec:affiliated}

Our first main result demonstrates that in a wide class of models, measuring a new covariate leads to a systematic shift in average effort across the population.  Further, the direction of this shift depends solely on whether the covariate is an attribute or a circumstance: Attributes reduce average effort, while circumstances increase it.

Our result applies to covariates that satisfy the following condition:

\begin{definition}[Affiliation] \label{ass:affiliation} 
Attribute 1 is \emph{Affiliated} if $(\theta_1, \eta)$ are statistically affiliated.\footnote{A pair of random variables $(Z_1, Z_2)$ is \emph{statistically affiliated} if $Z_2 \mid Z_1$ satisfies the monotone likelihood ratio property with respect to $Z_1.$  When $(Z_1, Z_2)$ possess a strictly positive, twice-differentiable joint density function $\rho(z_1, z_2)$, this condition is equivalent to $\partial^2 \log \rho/\partial z_1 \partial z_2 \geq 0$ everywhere.}  Circumstance 1 is \emph{Affiliated} if $(\eps_1, \delta)$ are statistically affiliated.
\end{definition}

Affiliation describes settings in which a good realization of a measured covariate (e.g., $\theta_1$) also implies good realizations (on average) of unmeasured covariates (e.g., $\eta$).  Examples of settings in which attribute 1 is Affiliated include: 

\begin{example} $(\theta_1, ..., \theta_J)$ follow a multivariate normal distribution, and all correlation coefficients are non-negative.
\end{example}

\begin{example}
$(\theta_1, ..., \theta_J)$ are iid draws from an exponential distribution with rate parameter $\lambda,$ where $\lambda \sim \text{Gamma}(\alpha_0, \beta_0)$ with $\alpha_0 \geq 1.$
\label{ex:Affiliation}
\end{example}

The following result establishes that a newly measured Affiliated attribute lowers average effort, while a newly measured Affiliated circumstance raises it.\footnote{\label{fn:strictMonotonicityAffiliated} We formally establish weak monotonicity. At the cost of a more involved proof,  It can further be shown  that when a covariate is non-degenerate, Affiliation implies strict monotonicity.}   

\begin{theorem} \label{thm:gencorr} Suppose Assumptions \ref{ass:regularity}-\ref{ass:expectationDiff} hold. If attribute $1$ is Affiliated, then measuring it weakly reduces average effort.  If circumstance $1$ is Affiliated, then measuring it weakly increases average effort.
\end{theorem}

To understand the result, consider  the simplest setting in which all covariates are independent of one another. (Independence is a sufficient condition for attribute 1 and circumstance 1 to be affiliated.) In this case, measuring a new attribute reduces the market's ex-post uncertainty about the agent's type. As a result, the market infers less about the agent's type from the outcome, decreasing the marginal value of improving the outcome through costly effort and therefore also the equilibrium level of effort. Measuring a new circumstance instead reduces the market's ex post uncertainty about the agent's shock, leading it to infer more about the agent's type from the outcome and boosting the agent's equilibrium effort.

When covariates are correlated,  measuring a new covariate need not reduce the market's ex-post uncertainty about all agents in the population. (See Section \ref{subsec:welfare} for an example.) To sign the average change in effort, we must show that even if uncertainty about some agents increases, their effort does not change by so much as to outweigh the opposing effort changes by all other agents. Affiliation ensures enough comovement of covariates to guarantee this result.

Theorem \ref{thm:gencorr} echos a similar finding by DJT in a related setting where the market observes an auxiliary signal only after the agent exerts effort.  (See their Example 5.3.)  Both results employ the same basic logic, but they rely on different statistical conditions. Our condition is formulated for our additive setting and highlights the role played by correlation between covariates, while DJT derive a more abstract condition in a setting with less structure on signals. Additionally, DJT assume technical regularity conditions that we are able to relax through an alternative proof technique.\footnote{Specifically, DJT implicitly assume sufficient regularity of the distributions of all random variables to permit two successive exchanges of an effort derivative and an expectation (see their Proposition 2.2). Our proof avoids one of these exchanges, and as a result we require only the weaker regularity condition imposed in Assumption \ref{ass:expectationDiff}.}  Finally, in our setting the change in average effort induced by a new measurement is only one aspect of its impact on the population distribution of effort. Our next results explore these implications in more depth.

\subsection{Redistribution of Uncertainty}\label{subsec:SH}

We now show that \emph{redistribution of uncertainty} across agents is necessary for measurement of a new covariate to impact higher moments of the population effort distribution.  Formally, whenever a covariate  reduces uncertainty about the corresponding outcome component in the same way for all agents, the average effect identified in Theorem \ref{thm:gencorr} is the sole effect of a new measurement.

The condition we identify requires that the variance as well as all higher moments of the unmeasured component are independent of the covariate realization. It strengthens the notion of homoskedasticity commonly imposed in linear regression models, which requires independence only of the residual variance. By analogy, we refer to it as \emph{Strong Homoskedasticity}.    

\begin{definition}[Strong Homoskedasticity]\label{def:homoskedasticity} Attribute 1 satisfies \emph{Strong Homoskedasticity} if the random variable $\eta - \mathbb{E}(\eta \mid a_1)$ is  independent of $a_1$.  Circumstance 1 satisfies \emph{Strong Homoskedasticity} if the random variable $\delta - \mathbb{E}(\delta \mid c_1)$ is  independent of $c_1$.
\end{definition}

Strong Homoskedasticity does not eliminate correlation between covariates, since the expectations of $\eta$ and $\delta$ are allowed to depend (respectively) on $a_1$ and $c_1.$ It does, however, rule out the possibility that the spread of an unmeasured covariate may depend on the measured covariate. To give a concrete example, suppose the measured covariate ``zip code" is correlated with the unmeasured covariate ``income." Strong Homoskedasticity allows the average income level to differ across zip codes, but is violated if income levels are more heterogeneous in some zip codes than in others.

Additional examples of Strongly Homoskedastic models include:

\begin{example} \label{ex:gaussian} $(\theta_1, \dots, \theta_J)$ and $(\eps_1, \dots, \eps_K)$ each follow multivariate normal distributions.
\end{example}

\begin{example} The type and shock components satisfy $\theta_j = F_j(\theta_1) + X_j$ for $j \geq 2$ and $\eps_k = G_k(\eps_1) + Z_k$ for $k \geq 2,$ for functions $F_j$ and $G_k$ and random variables $X_j \indep \theta_1$ and $Z_k \indep \eps_1$.
\end{example}

Under Strong Homoskedasticity, a strong version of Theorem \ref{thm:gencorr} ensures a uniform change of effort across agents.\footnote{\label{fn:strictMonotonicitySH} As in Theorem \ref{thm:gencorr}, we formally establish only weak monotonicity.  When a covariate is additionally non-degenerate, Strong Homoskedasticity further implies strict monotonicity.  (See footnote \ref{fn:strictMonotonicityAffiliated}.)}  Since Strong Homoskedasticity encompasses correlation structures outside the class of affiliated covariates (for instance, multivariate normal models with negative correlation coefficients), this result further broadens the finding of Theorem \ref{thm:gencorr}.

\begin{theorem} \label{thm:mean-shift}
Suppose Assumptions \ref{ass:regularity}-\ref{ass:expectationDiff} hold. If attribute 1 satisfies Strong Homoskedasticity, then measuring it weakly reduces every agent's effort.  If circumstance 1 satisfies Strong Homoskedasticity, then measuring it weakly increases every agent's effort. 

In either case, the magnitude of the effort change is the same for every agent, irrespective of their covariate realization.
\end{theorem}

When Strong Homoskedasticity fails, measurement of a new covariate induces heterogeneity in the spread of the market's beliefs about different agents, an effect we refer to as redistribution of uncertainty.  This redistribution can lead to heterogeneous effort changes across the population, and may even exert a directional effect on effort which for some agents overturns the average effect identified in Theorem \ref{thm:gencorr}.  We provide an illustration of this possibility through an example in Section \ref{subsec:welfare}.

 Our results therefore highlight an important connection between redistribution of uncertainty and inequality generated by data usage. When a new measurement redistributes uncertainty across agents, its payoff impact may differ across agents with different covariate realizations, leading some agents to become better off at the expense of others.\footnote{If agents do not receive any up-front payments in period 1, they become better or worse off to the extent their effort falls or rises.  In some contexts, agents might be compensated for the value generated by their equilibrium effort.  (See, for instance, the competitive labor market of \citet{Holmstrom}.)  In that case, agents' payoffs increase in effort over some range.  Nonetheless, redistribution of uncertainty still benefits some agents at the expense of others.} A social planner may care intrinsically about this disparate impact, especially since these payoff changes may be correlated with sensitive social or demographic characteristics via the measured covariates.\footnote{Automated prediction algorithms have recently come under scrutiny for unintentionally discriminating against protected social groups such as racial minorities on the basis of such correlations \citep{EthicalAlg}.  Our analysis highlights a new channel through which such correlations might harm disadvantaged groups.} In the next section we show that even when the social planner does not have any intrinsic distributional concerns, this dispersion harms aggregate welfare.

\subsection{Social Welfare and Data Regulation}\label{subsec:welfare}

An important question faced by regulators is whether to allow firms in particular markets to use new covariates for forecasting. We now apply our positive results to answer this normative question. 
We characterize the welfare impact of measuring \emph{regular covariates}, a class encompassing all covariates exhibiting the systematic impact on average effort identified in Theorems \ref{thm:gencorr} and \ref{thm:mean-shift}.\footnote{To facilitate non-quadratic effort costs (see Section \ref{subsec:genCost}), we define regularity with respect to the marginal value of effort rather than effort itself.  Since equilibrium effort is proportional to the marginal value of effort when costs are quadratic, the two notions are equivalent in our baseline setting.}

\begin{definition} \label{def:regularity}
Attribute 1 is \emph{regular} if measuring it weakly reduces the marginal value of effort on average.  Circumstance 1 is \emph{regular} if measuring it weakly increases the marginal value of effort on average.  If the change is strict, we call the covariate \emph{strictly} regular.
\end{definition}

\noindent It follows from Theorems \ref{thm:gencorr} and \ref{thm:mean-shift} that all Affiliated and Strongly Homoskedastic covariates are regular.\footnote{As noted in footnotes \ref{fn:strictMonotonicityAffiliated} and \ref{fn:strictMonotonicitySH}, when a covariate is additionally nondegenerate, it can be shown to be strictly regular.}

\begin{figure}[ht]
    \centering
    \includegraphics[scale=0.5]{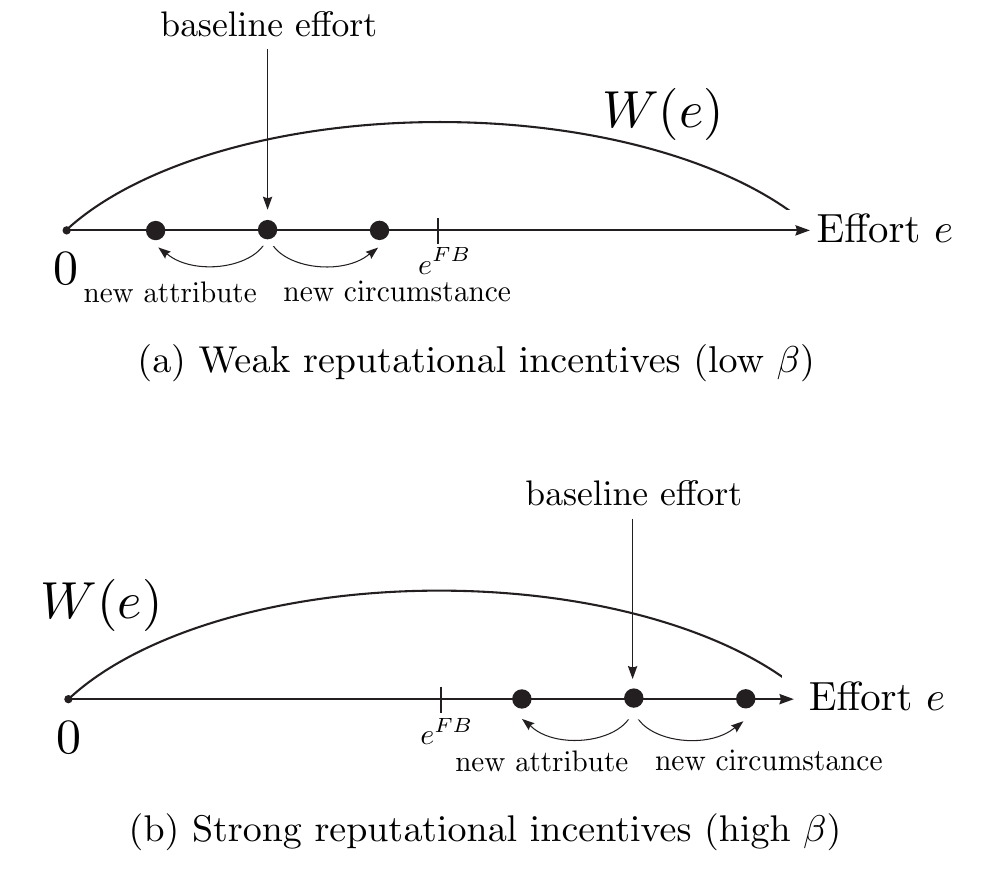}
    \caption{\footnotesize{The welfare impact of measuring a new covariate when effort is deterministic.}}
    \label{fig:welfare}
\end{figure}

To develop some intuition for the impact of a new measurement on welfare, suppose first that effort in both the baseline and expanded environments is deterministic, for example because the newly measured covariate is Strongly Homoskedastic. Let
$MV \equiv MV_{\emptyset,\emptyset}$ and $MV_{+A} \equiv MV_{\{1\},\emptyset}$ denote  the marginal value of effort before and after attribute 1 is measured.  We will similarly let $e$ and $e_{+A}$ denote effort in these two environments.  The agent's effort is related to his marginal value of effort via
\[e = \frac{\beta}{1 - \beta} \cdot MV, \quad e_{+A} = \frac{\beta}{1 - \beta} \cdot MV_{+A}.\]
If attribute 1 is regular, then $MV_{+A} \leq MV$ and so correspondingly $e_{+A} \leq e$. Whether this change is welfare-improving depends on the size of $\beta$.   If $\beta$ is sufficiently large, then measuring the attribute moves the agent's effort closer to the first-best level $e^{FB}=1$, increasing welfare; in contrast, if $\beta$ is sufficiently small, then effort moves away from first-best and welfare is reduced. This logic is reversed for circumstances. See Figure \ref{fig:welfare} for an illustration.

When agents react heterogeneously to a newly measured covariate, aggregate welfare depends on details of the distribution of effort beyond the mean.  Signing the effect of a covariate on welfare therefore requires aggregating the systematic effort effect identified above and the additional impact of redistributing uncertainty.   

The following theorem shows that even in the presence of disparate impact, the effect of a newly measured covariate on aggregate welfare depends in a simple way on the size of the reputation weight.
Specifically, for every regular attribute, there is a threshold reputation weight $\beta^*$ such that measuring the covariate is welfare-improving only for reputation weights above $\beta^*$.\footnote{We say that a covariate is welfare-improving if aggregate welfare \emph{strictly} increases when the covariate is measured.} Analogously, for every regular circumstance, there is a threshold reputation weight $\beta_*$ such that measuring the covariate is welfare-improving only for all reputation weights below $\beta_*$. 

\begin{theorem}\label{thm:welfare}
 Suppose attribute 1 is regular. Then there exists a threshold reputation weight $\beta^* \in (0, 1]$ such that measuring the attribute is welfare-improving if and only if $\beta > \beta^*.$  Moreover, $\beta^* <1$ if and only if \begin{equation} \label{eq:squared}
\mathbb{E}\left(MV_{+A}^2\right) < \mathbb{E}\left(MV^2\right).
\end{equation}

Suppose circumstance 1 is regular. Then there exists a threshold reputation weight $\beta_* \in [0, 1)$ such that measuring the circumstance is welfare-improving if and only if $\beta < \beta_*.$  Moreover, $\beta_* > 0$ if and only if the circumstance is strictly regular.
\end{theorem}

Despite the apparent symmetry between the two parts of this result, they are not perfectly mirrored. Every (strictly) regular circumstance is welfare-improving for sufficiently small $\beta$, but some attributes fail to improve welfare even for large $\beta$.  This asymmetry stems from the fact that a larger reputation weight leads to increased effort dispersion in the presence of uncertainty redistribution, and hence increased aggregate effort costs.  For attributes, this effect is largest precisely when the average effort decrease from measuring the covariate is most beneficial to welfare, leading to an ambiguous relationship between reputational concerns and total welfare.  By contrast, for circumstances the dispersion effect is most pronounced when the effort increase from measurement is most harmful to welfare, and so the two forces reinforce one another.

Condition \eqref{eq:squared} characterizes when the dispersion effect overwhelms the average effect of measuring a new attribute. If the attribute decreases the marginal value of effort for all agents, then a fall in the expected marginal value of effort implies that the expectation of its square falls as well, satisfying \eqref{eq:squared}.  The theorem then guarantees that the attribute improves welfare for all sufficiently large reputation weights.  By contrast, when the newly measured attribute increases effort for some agents and decreases it for others, the dispersion of the marginal value of effort $MV_{+A}$ generated by redistribution of uncertainty may be sufficiently large that(\ref{eq:squared}) fails.  In this case, measuring the attribute diminishes welfare regardless of the weight agents place on their future reputation. 

This result highlights the role that redistribution of uncertainty plays in determining welfare.  Even if measuring a new covariate leads to a welfare-improving shift in aggregate effort, it may be optimal to prohibit use of this covariate for forecasting if the aggregate shift is achieved at the cost of large welfare losses associated with particular groups of agents. We conclude our analysis with a simple example demonstrating this possibility.

\paragraph{Example.} Worker productivity $\theta$ is a function of two attributes: residential stability $a_1 \in [0,1]$ and worker reliability $a_2 \in [0,1]$. These attributes determine the worker's productivity via 
$\theta = f^1(a_1) + f^2(a_2),$ 
where
\begin{align*}
    f^1(a_1) = a_1/10, \quad \quad \quad
    f^2(a_2) = 10 \cdot a_2
\end{align*}
There are no measurable circumstances, and the total shock to output $\eps$ follows the standard normal distribution $\mathcal{N}(0, 1).$ 

Residential stability and worker reliability are positively correlated, with residential stability $a_1$ distributed uniformly on $[0,1]$, while the conditional distribution of worker reliability $a_2$ given $a_1$ is:
\[a_2 \mid a_1 \sim \begin{cases}
U([0.9, 1]), & \forall a_1 \geq 0.05 \\
U([0, 1]), & \forall a_1 < 0.05 \end{cases} \]
That is, workers with very low residential stability (bottom 5\%) are less reliable on average, but are also substantially more heterogeneous.\footnote{For example, some workers may move frequently due to evictions, reflecting low reliability.  On the other hand, the exceptionally prolific mathematician Paul Erd\H{o}s famously possessed no permanent residence (\url{https://www.scientificamerican.com/article/an-arbitrary-number-of-years-since-mathematicians-birth/}).}  

In a labor market in which neither attribute is measured, the market learns about worker productivity solely through output.  All workers then exert a common level of effort, which can be numerically computed to be $e^* \approx 0.16 \cdot \frac{\beta}{1 - \beta}$. If the market begins measuring residential stability, equilibrium effort for workers with high residential stability $(a_1 > 0.05)$ falls by roughly 50\%, to  $e^{**}_H \approx 0.077 \cdot \frac{\beta}{1 - \beta},$ while effort for all remaining workers rises fivefold to $e^{**}_L \approx 0.82 \cdot \frac{\beta}{1 - \beta}$. Intuitively, although residential stability has only a small direct impact on job performance, it has a large impact on the market's uncertainty about worker reliability. The market's uncertainty about workers who move very frequently (residential stability falls below the bottom 5\% percentile) \emph{increases}, while the market's uncertainty about the type of all remaining workers decreases. Since, moreover,
\[\mathbb{E}\left(MV_{+A}^2\right) \approx 0.039 > MV^2 \approx 0.026,\]
Theorem \ref{thm:welfare} implies that aggregate welfare decreases upon measuring attribute 1 for any reputation weight $\beta.$

\section{Extensions}\label{sec:extensions}

We now analyze several extensions of our framework. Section \ref{subsec:correlation} studies the effect of covariates which are correlated with both the type and shock.  Section \ref{subsec:modelUncertainty} establishes that our results are robust to agent uncertainty about the market's beliefs.  Section \ref{subsec:genCost} relaxes the assumption that the agent's effort costs are quadratic. 

\subsection{General Covariates}\label{subsec:correlation}
Our main results have assumed that individual covariates are informative about the agent's type $\theta$ or shock $\eps,$ but not both.  In some applications, covariates may plausibly lie somewhere between these two extremes.  We now show how our results can be adapted to accommodate such covariates.

As in our main results, we focus on a baseline in which no covariates are observed and an expanded dataset consisting of a single covariate, whose value we will denote by the random variable $X.$  We allow this covariate to be correlated with both $\theta$ and $\eps$ in a very general way, which we summarize by its effect on the conditional mean of the outcome.  Let $Y^0 \equiv \theta + \eps$ be the baseline outcome ignoring the agent's effort, and define the random variable $\bar{Y}^0 \equiv \mathbb{E}(\theta + \eps \mid X)$ to be the conditional mean of the baseline outcome given the covariate.  We maintain an invertibility assumption ensuring that measuring $X$ is equivalent to observing the conditional mean outcome $\bar{Y}^0:$
\begin{assumption}[General invertibility]\label{ass:invertibilityCorrelated}
$\mathbb{E}(\theta + \eps \mid X = x)$ is a one-to-one function of $x.$
\end{assumption}\label{ass:admissibilityCorrelated}
\noindent This condition is analogous to the invertibility assumptions impose on the effect size functions $f^j$ and $g^k$ in the baseline model, and it serves the same purpose.

We additionally impose admissibility and differentiability assumptions analogous to Assumptions \ref{ass:regularity} and \ref{ass:expectationDiff} in the baseline analysis.
\begin{assumption}[General admissibility]\label{ass:regularityCorrelated}
$(\bar{Y}^0, Y^0)$ are statistically affiliated.
\end{assumption}

\begin{assumption}[General differentiability]\label{ass:expectationDiffCorrelated}
For every effort level $e,$ the derivative $\frac{\partial}{\partial Y}\mathbb{E}^e\left(\theta \mid Y\right)$ exists and is uniformly bounded across all realizations of $Y.$  For every effort level $e$ and realization of $X,$ the derivative
$\frac{\partial}{\partial Y}\mathbb{E}^e\left(\theta \mid Y, X\right)$ exists and is uniformly bounded across all realizations of $Y$.
\end{assumption}
Assumption \ref{ass:regularityCorrelated} serves the same role as Assumption \ref{ass:regularity} in ensuring that better outcomes correspond to improved inferences about latent variables.  In particular, if the ``net residual'' $\Delta Y^0 \equiv Y^0 - \bar{Y}^0$ is independent of the realization of $X,$ affiliation of $(\bar{Y}^0, Y^0)$ reduces to log-concavity of the density function of $\Delta Y^0.$ Meanwhile, Assumption \ref{ass:expectationDiffCorrelated} is a straightforward adaptation of Assumption \ref{ass:expectationDiff}. 

Finally, we impose an assumption ensuring that $\theta$ and $\eps$ are correlated only through the covariate $X.$  We maintain it to focus on the simplest context in which correlation between the type and shock might arise.  
\begin{assumption}[General independence]\label{ass:conditionalIndependence}
$(\theta, \eps)$ are independent conditional on $X$.
\end{assumption}

We now derive conditions under which measuring $X$ increases or decreases effort, extending the results of Theorem \ref{thm:gencorr} to this setting.

\begin{proposition}\label{prop:MVDiffCorr}
Suppose that Assumptions \ref{ass:invertibilityCorrelated}-\ref{ass:conditionalIndependence} hold.
\begin{enumerate}
    \item If $(\bar{Y}^0, \theta)$ and $(\bar{Y}^0, -\eps)$ are each statistically affiliated, then measuring $X$ reduces average effort.
    
        \item If $(\bar{Y}^0, -\theta)$ and $(\bar{Y}^0, \eps)$ are each statistically affiliated, then measuring $X$ increases average effort.
\end{enumerate} 
\end{proposition}

This result establishes that a covariate which is positively associated with one component of the outcome, and is simultaneously negatively associated with the remaining component, has an unambiguous impact on the expected marginal value of effort.  The positive association condition here is a direct analog of the affiliation condition in Theorem \ref{thm:gencorr}, and is needed for the same reason.  Meanwhile, the negative association condition rules out scenarios in which a good covariate realization implies both a high type \emph{and} a high shock.  Since these inferences have conflicting effects on the marginal value of effort, the net effect of measuring such a covariate is inherently ambiguous.  By contrast, if a good covariate realization suggests a high type and a low shock, or vice versa, the two effects reinforce and the measurement has an unambiguous impact on average effort.

This result can be strengthened to obtain a uniform effect on effort under homoskedasticity conditions similar to those imposed in Theorem \ref{thm:mean-shift}.  In particular, let $\Delta \theta \equiv \theta - \mathbb{E}(\theta \mid X)$ be the residual unobserved type component after measuring $X.$  Define $\Delta \eps$ similarly with respect to the shock.  Then if the joint distribution of $(\Delta \theta, \Delta \eps)$ is independent of the realization of $X,$ measuring $X$ affects effort uniformly across all agents.  

To illustrate these forces concretely, we analyze the effect of measuring a new covariate in a multivariate Gaussian setting.  Suppose that $\theta$ and $\eps$ are decomposable as
\[\theta = \mu + b \cdot X + Z, \quad \eps = d \cdot X + W \]
where $X \sim N(0, \sigma_x^2)$, $Z \sim N(0, \sigma_z^2)$, and $W \sim N(0, \sigma_w^2)$ are mutually independent and $\mu,$ $b$, and $d$ are known constants.  The following lemma ensures that the regularity assumptions imposed in Proposition \ref{prop:MVDiffCorr} are satisfied in this setting whenever $b + d \neq 0$, a condition we will maintain going forward.\footnote{If $b + d = 0,$ then $X$ does not impact $Y$ and cannot be estimated by observing the outcome.  As a result, measuring it has no impact on the marginal value of effort.}.

\begin{lemma}\label{lemma:regularityGaussian}
Assumptions \ref{ass:invertibilityCorrelated} through \ref{ass:conditionalIndependence} are satisfied in a multivariate Gaussian setting with a general covariate whenever $b + d \neq 0$.
\end{lemma}

We now check when the conditions identified in Proposition \ref{prop:MVDiffCorr} under which measuring $X$ increases or decreases effort are satisfied.  $(\bar{Y}^0, \theta)$ and $(\bar{Y}^0, \eps)$ are each jointly Gaussian, and $(\bar{Y}^0, \pm \theta)$ are positively correlated iff $\text{sign}(b + d) = \pm \text{sign}(b).$  Similarly, $(\bar{Y}^0, \pm \eps)$ are positively correlated iff $\text{sign}(b + d) = \pm \text{sign}(d).$  Then whenever $b > 0,$ Proposition \ref{prop:MVDiffCorr} implies that measuring $X$ reduces effort if $b + d > 0 \geq d,$ i.e., $d \in (-b, 0].$  Similarly, whenever $d > 0,$ measuring $X$ increases effort if $b + d > 0 \geq b,$ i.e., $b \in (-d, 0].$  

The bounds $d \leq 0$ and $b \leq 0$ illustrate the general point made earlier: Measuring $X$ has an unambiguous effect on effort only if its informativeness about one component of the outcome is reinforced rather than opposed by its informativeness about the remaining component.  The remaining condition $b + d > 0$ ensures that better outcomes correspond to improved inferences about the type or shock in the baseline, without which the expected directional effect of a measurement can reverse.

We can verify these results by explicitly calculating $MV$ and $MV_+$, the marginal value of effort before and after measuring $X.$  The following result summarizes the calculation.

\begin{proposition}\label{prop:correlatedGaussian}
$\text{sign}\left(MV - MV_+\right) = \text{sign}\left((b+d)\left(\frac{b}{\sigma_z^2} - \frac{d}{\sigma_w^2}\right)\right).$
\end{proposition}

If $b > 0$ and $d \in (-b, 0],$ this result implies that $MV_+ < MV,$ in line with the prediction of Proposition \ref{prop:MVDiffCorr}.  Similarly, if $d > 0$ and $b \in (-d, 0],$ then $MV_+ > MV$.  Conversely, if both $b$ and $d$ are positive, the sign of $MV - MV_+$ is ambiguous.  Depending on the sizes of these coefficients relative to the residual uncertainty about $\theta$ and $\eps,$ measuring $X$ could move the marginal value of effort in either direction.

\subsection{Model Uncertainty and Misspecification} \label{subsec:modelUncertainty}

Suppose that, contrary to our assumptions in the baseline model, the agent is subjectively uncertain about the market's perceived distribution of $(\theta,\eps)$ given the agent's measured covariates. Such a situation may arise if he does not know which set of covariates the market observes, or if he does not know how the market maps his covariate values into perceived type and shock distributions. 

This subjective uncertainty can be modeled by supposing the agent possesses beliefs over possible joint distributions of $(\theta,\eps)$ that the market might hold when forecasting the agent's type. (It is not important that the market's true model be contained in the support of the agent's beliefs, so the agent may be misspecified.)  We will continue to maintain the assumption that the agent is not asymmetrically informed about his type, and so his own subjective belief about the distribution of his type and outcome is the expectation of his belief about the market's distribution.

In this setting, all of our results extend in the following sense: If the agent becomes convinced that the market's statistical model has become ``better-informed'' about the agent's type or shock, his effort will move in the direction predicted by our results, so long as the corresponding statistical assumptions hold for each model in the support of the agent's beliefs.  More precisely, an agent believes the market has become ``better-informed'' if he thinks that, regardless of what statistical model it is in fact using, the market has gained access to an additional attribute or additional covariate.  In that case, the marginal value of effort moves in the same direction conditional on any model in the support of the agent's beliefs, and therefore the expected marginal value of effort moves in this direction as well.  Our main results therefore continue to hold in this environment.

\subsection{General Convex Cost Functions} \label{subsec:genCost}
We have established our main results under the assumption that effort costs take the form $C(e) = \frac{1}{2} e^2.$  Under this cost function, equilibrium effort is identical to the marginal value of effort, allowing us to characterize the former by analyzing the latter.  More generally, when $C$ is a strictly convex cost function, equilibrium effort is a uniquely determined, strictly increasing function of the marginal value of effort:
\[e^*_{\mathcal{J},\mathcal{K}} = (C')^{-1}\left(\frac{\beta}{1 - \beta} \cdot MV_{\mathcal{J},\mathcal{K}}\right),\]
where $MV_{\mathcal{J},\mathcal{K}}$ is  as defined in (\ref{def:MVexpand}).  As a result, under such a cost function, a deterministic shift in the marginal value of effort implies a change in effort in the same direction.  This implies in particular that the results of Theorem \ref{thm:mean-shift} under Strong Homoskedasticity extend immediately. 

Theorem \ref{thm:gencorr} for Affiliated covariates extends so long as all agents change their effort in the same direction, and more generally under a condition on the third derivative of the effort cost function.\footnote{The proof of this result is a straightforward application of the proof of Theorem \ref{thm:gencorr}, combined with the logic of the discussion following the proposition statement.} (In Appendix \ref{app:WelfareCost} we present similar, but more restrictive, generalizations of the welfare results from Section \ref{subsec:welfare}.)
\begin{proposition}\label{prop:gencorrNonQuadratic}
Suppose Assumptions \ref{ass:regularity}-\ref{ass:expectationDiff} hold.
\begin{enumerate}[(a)]
\item  If attribute $1$ is Affiliated, then measuring it reduces average effort if either $C''' \geq 0$ or else all agents change their effort in the same direction.  

\item If circumstance $1$ is Affiliated, then measuring it increases average effort if either $C''' \leq 0$ or else all agents change their effort in the same direction.
\end{enumerate}
\end{proposition}
The new force which arises under general cost functions is that average effort may respond to mean-preserving spreads of the marginal value of effort.  To illustrate the idea, consider any cost function $C(e) \propto e^k,$ where $k > 1.$  If $k > 2,$ then under such a cost function the marginal cost of effort is convex, so equilibrium effort is a concave function of the marginal value of effort.  Hence any mean-preserving spread of the marginal value of effort reduces average effort.  Conversely, if $2 > k > 1,$ effort is a convex function of the marginal value of effort, and a mean-preserving spread of the marginal value of effort increases average effort.

Measuring a new Affiliated covariate has two effects: It shifts the \emph{average} marginal value of effort, and (whenever Strong Homoskedasticity fails) it may additionally introduce a \emph{spread} in the distribution of marginal values.  If the marginal cost of effort is convex, this second effect tends to reduce equilibrium effort.  Thus when a new attribute is measured, these two forces work together to lower average effort, and the results of Theorem \ref{thm:gencorr} continue to hold.  A similar outcome holds when the marginal cost of effort is concave and a new circumstance is measured. When the two forces conflict, the net effect on effort can be ambiguous.  In particular, if agents change their effort in different directions, average effort could move in the opposite direction from the average marginal value of effort.  

\section{Conclusion}\label{sec:conclusion}

As firms and governments move towards collecting large consumer datasets as inputs to decision-making, the question of whether and how to regulate the usage of personal data has emerged as an important policy question.  Recent regulations, such as the European Union's General Data Protection Regulation, have focused on protecting consumer privacy and improving transparency regarding what kind of data is being collected. An important complementary consideration is how data impacts economic outcomes.  In this paper we have focused on one such factor\textemdash  the effect that market access to novel covariates has on incentives for hidden effort.  

Our results indicate that forecasting from data on enduring personal attributes decreases average effort across the population, while conversely data reflecting short-lived circumstances boosts effort.  It is therefore important to distinguish between these two classes of data when regulating data usage. Further, new data may lead to increased variation in effort across workers, an outcome which has a first-order impact on welfare.  This finding suggests that regulators should also take into account the distributional effects of new data when deciding whether to permit its use in particular markets.

One way to interpret the attributes and circumstances in our model is as stand-ins for covariates with different levels of persistence in a dynamic model, where the agent exerts effort over multiple periods and his type evolves over time.  Generalizing our results
to a many-period setting is technically challenging due to the possibility of belief divergences following effort deviations under non-Gaussian information structures.  Nonetheless, doing so would permit a richer study of the welfare implications of forecasting from data with varying persistence, making it an important avenue for future research.

\pagebreak

\begin{center}
    \huge{Appendix}
\end{center}

\appendix

\section{Alternative Welfare Specifications}\label{app:welfareAlt}

In this appendix we extend our welfare analysis to consider alternative environments in which effort improves future as well as current outcomes (``learning-by-doing") or is partially dissipative  (``gaming" effort).

\subsection{Learning-by-Doing} 

In some applications, effort may improve future as well as current outcomes, for instance in labor market settings featuring learning-by-doing.  In that case, the agent's type is not constant over time but instead improves with past effort, and effort has socially beneficial effects in multiple periods.  

Our model can be modified to accommodate this feature by allowing the agent's type $\theta^{(t)}$, which determines the average outcome in period $t,$ to be time-dependent.  Concretely, we will suppose that $\theta^{(2)} = \theta^{(1)} + \gamma \cdot e,$ where $\gamma > 0$ is a learning-by-doing parameter.  Period-1 output is
\[Y = e + \theta^{(1)} + \eps,\]
while the agent's period-2 reputational reward is $\mathbb{E}(\theta^{(2)} \mid Y).$  

The presence of learning by doing does not affect equilibrium effort, because the agent's reputational reward is based on the market's forecast of his effort (which is fixed) rather than his true effort. This expectation is
\[\mathbb{E}^{e^*}(\theta^{(2)} \mid Y) = (1+\gamma) \cdot e^* + \mathbb{E}^{e^*}(\theta^{(1)} \mid Y).\]
Exerting additional effort is therefore valuable to the agent only insofar as it improves the market's forecast of $\theta^{(1)},$ exactly as in our main model. Thus equation \eqref{eq:EffortRandom} continues to characterize equilibrium effort. 

The socially optimal effort level, however, becomes $e^{FB} = 1 + \gamma$ in this model. Equilibrium effort therefore falls below the first-best level for a broader range of reputation weights $\beta$ as the learning-by-doing parameter $\gamma$ increases.  An analogue of Theorem \ref{thm:welfare} continues to hold, where the threshold reputation weights $\beta^*$ and $\beta_*$ are increasing in $\gamma.$  In other words, increased learning-by-doing makes circumstances (which boost effort) more attractive and attributes (which reduce it) less so at any given reputational weight.

\subsection{``Gaming" Effort} 
In other applications, effort may be dissipative and serve to distort a signal of quality without producing social value.  This possibility may arise, for instance, in labor market settings in which a worker can spend time performing ``influence activities'' to increase the visibility of his accomplishments (as in \citet{influenceActivities}).  It may also arise in educational settings where the outcome variable is a test score that can be improved by test prep with no further educational value (as in \citet{FrankelKartik}).  

To accommodate this possibility, our welfare criterion can be modified to discount the welfare benefits of effort:
\[w(\theta, e) = \theta + \delta \cdot e - \frac{1}{2} e^2,\]
where $\delta \in [0, 1)$ measures the proportion of effort which is socially beneficial.  When $\delta = 0,$ effort is totally unproductive, while $\delta \in (0, 1)$ captures situations in which some fraction of effort contributes social value.

The dissipative nature of effort has no impact on equilibrium effort, but reduces the first-best effort level to $e^{FB} = \delta.$  Equilibrium effort will therefore exceed the first-best level for a broader range of reputation weights $\beta$ as effort becomes increasingly dissipative.  A result analogous to Theorem \ref{thm:welfare} can be established in this setting, with the threshold reputation weights $\beta^*$ and $\beta_*$ increasing in $\delta.$  One interesting case is $\delta = 0$, in which effort is fully dissipative effort.  In that case, measuring new attributes improves welfare while measuring new circumstances diminishes it, regardless of the reputation weight $\beta$.  (The one exception is for an attribute with significant disparate impact, which may still be welfare-reducing for all $\beta.$)

\section{Results for a General Baseline} \label{app:GenBaseline}

The results of Section  \ref{sec:results} can be straightforwardly generalized to accommodate settings in which some covariates are initially measured by the market.  Given any sets $\mathcal{J} \subseteq \{1, \dots, J\}$ of measured attributes and $\mathcal{K} \subseteq \{1, \dots, K\}$ of measured circumstances, define
\[
\eta_{\mathcal{J}} \equiv \sum_{j \notin \mathcal{J}} \theta_j + u_\theta, \quad \delta_\mathcal{K} \equiv \sum_{k \notin \mathcal{K}} \eps_k + u_\eps
\]
to be the sums of all \emph{unmeasured} components of the agent's type and shock.

Fix a baseline family $(\mathcal{J}, \mathcal{K})$ of measured covariates.  Affiliation and Strong Homoskedasticity may be generalized to this environment as follows:

\begin{definition}[Affiliation]
The attribute $j' \notin \mathcal{J}$ is $\mathcal{J}$-\emph{Affiliated} if $(\theta_{j'}, \eta_{\mathcal{J} \cup \{j'\}})$ is affiliated conditional on $a_{\mathcal{J}}.$  The circumstance $k' \notin \mathcal{K}$ is $\mathcal{K}$-\emph{Affiliated} if $(\eps_{k'}, \delta_{\mathcal{K} \cup \{k'\}})$ is affiliated conditional on $c_{\mathcal{K}}.$
\end{definition}

\begin{definition}[Strong Homoskedasticity]
The attribute $j' \notin \mathcal{J}$ satisfies $\mathcal{J}$-\emph{Strong Homoskedasticity} if $\eta_{\mathcal{J} \cup \{j'\}} - \mathbb{E}(\eta_{\mathcal{J} \cup \{j'\}} \mid a_{\mathcal{J} \cup \{j'\}})$ is independent of $a_{j'}$ conditional on $a_{\mathcal{J}}$.  The circumstance $k' \notin \mathcal{K}$ satisfies $\mathcal{K}$-\emph{Strong Homoskedasticity} if $\delta_{\mathcal{K} \cup \{k'\}} - \mathbb{E}(\delta_{\mathcal{K} \cup \{k'\}} \mid c_{\mathcal{K} \cup \{k'\}})$ is independent of $c_{k'}$ conditional on $c_{\mathcal{K}}.$
\end{definition}

As formulated, these definitions apply across all $(\mathcal{J},\mathcal{K})$-\emph{subpopulations} of agents, where each subpopulation consists of all agents sharing a particular realization of $(a_{\mathcal{J}}, c_{\mathcal{K}})$.  They could alternatively be formulated more narrowly to apply only for a particular set of realized covariates, if the analyst is primarily interested in the impact of a new covariate on a particular subpopulation of agents.

Assumptions \ref{ass:regularity} and \ref{ass:expectationDiff}, which imposed log-concavity on latent variables and boundedness of derivatives of conditional expectations, must also be extended for a general set of baseline measured covariates.  We split these conditions into an assumption we maintain in the baseline environment, and a condition imposed on newly measured covariates.

\begin{assumption}[Baseline Admissibility]\label{ass:admissibilityGeneralBaseline}
The conditional distributions $\theta \mid a_{\mathcal{J}}$ and $\eps \mid c_{\mathcal{K}}$ have log-concave density functions, and for every effort level $e$ and realization of $(a_{\mathcal{J}}, c_{\mathcal{K}}),$ the derivative
$\frac{\partial}{\partial Y}\mathbb{E}^e\left(\theta \mid Y, a_{\mathcal{J}}, c_{\mathcal{K}}\right)$
exists and is uniformly bounded across all realizations of $Y$.
\end{assumption}

\begin{definition}[Admissible covariates]
An attribute $j' \notin \mathcal{J}$ is $\mathcal{J}$-\emph{admissible} if $\eta_{\mathcal{J} \cup \{j'\}} \mid a_{\mathcal{J} \cup \{j'\}}$ has a log-concave density function for every realization of $a_{\mathcal{J} \cup \{j'\}},$ and for every effort level $e$ and realization of covariates $(a_{\mathcal{J} \cup \{j'\}}, c_{\mathcal{K}}),$ the derivative
$\frac{\partial}{\partial Y}\mathbb{E}^e\left(\theta \mid Y, a_{\mathcal{J} \cup \{j'\}}, c_{\mathcal{K}}\right)$
exists and is uniformly bounded across all realizations of $Y$.  

A circumstance $k' \notin \mathcal{K}$ is $\mathcal{K}$-\emph{admissible} if $\delta_{\mathcal{K} \cup \{k'\}} \mid c_{\mathcal{K} \cup \{k'\}}$ has a log-concave density function for every realization of $c_{\mathcal{K} \cup \{k'\}}$, and for every every effort level $e$ and realization of covariates $(a_{\mathcal{J}}, c_{\mathcal{K} \cup \{k'\}}),$ the derivative
$\frac{\partial}{\partial Y}\mathbb{E}^e\left(\theta \mid Y, a_{\mathcal{J}}, c_{\mathcal{K} \cup \{k'\}}\right)$
exists and is uniformly bounded across all realizations of $Y$.  
\end{definition}

With these concepts, we can generalize Theorems  \ref{thm:gencorr} and \ref{thm:mean-shift} as follows:

\begin{theorem} \label{thm:genBaseA} Suppose Assumption \ref{ass:admissibilityGeneralBaseline} holds.
\begin{itemize}
\item[(a)] If attribute $j'$ is $\mathcal{J}$-admissible and satisfies $\mathcal{J}$-Affiliation, then measuring it weakly reduces average effort within each $(\mathcal{J}, \mathcal{K})$-subpopulation.

\item[(b)] If circumstance $k'$ is $\mathcal{K}$-admissible and satisfies $\mathcal{K}$-Affiliation, then measuring it weakly increases average effort within each $(\mathcal{J}, \mathcal{K})$-subpopulation.
\end{itemize}

\end{theorem}

\begin{theorem} \label{thm:genBaseMS}
Suppose Assumption \ref{ass:admissibilityGeneralBaseline} holds.
\begin{itemize}
    \item[(a)] If attribute $j'$ is $\mathcal{J}$-admissible and satisfies $\mathcal{J}$-Strong Homoskedasticity, then measuring it weakly reduces every agent's effort.  Further, the magnitude of the effort change is the same for every agent in each $(\mathcal{J}, \mathcal{K})$-subpopulation.
\item[(b)] If circumstance $k'$ is $\mathcal{K}$-admissible and satisfies $\mathcal{K}$-Strong Homoskedasticity, then measuring it weakly increases every agent's effort.  Further, the magnitude of the effort change is the same for every agent in each $(\mathcal{J}, \mathcal{K})$-subpopulation.
\end{itemize}
\end{theorem}

These results can be applied repeatedly to assess the impact of measuring multiple covariates, so long as admissibility and the corresponding statistical condition (Affiliation or Strong Homoskedasticity) holds for each of the measured covariates relative to its respective baseline.  Note in particular that the exponential and multivariate normal settings of Examples \ref{ex:Affiliation} and \ref{ex:gaussian} satisfy Affiliation and Strong Homoskedasticity, respectively, for any baseline and newly measured covariate.  

Our welfare results also hold under general baselines using appropriate notions of regularity and strict regularity.

\begin{definition} \label{def:regularityGeneralBaseline}
Fix a baseline family of measured covariates $(\mathcal{J}, \mathcal{K}).$  Then:
\begin{itemize}
    \item An attribute $j' \notin \mathcal{J}$ is $(\mathcal{J},\mathcal{K})$-\emph{regular} if, conditional on any realization of $(a_\mathcal{J}, c_\mathcal{K})$, measuring $j'$ weakly reduces the marginal value of effort on average.  It is \emph{strictly} $(\mathcal{J},\mathcal{K})$-regular if the reduction is strict for a positive fraction of realizations of $(a_\mathcal{J}, c_\mathcal{K})$.
    
    \item A circumstance $k' \notin \mathcal{K}$ is $(\mathcal{J},\mathcal{K})$-\emph{regular} if, conditional on any realization of $(a_\mathcal{J}, c_\mathcal{K})$, measuring $k'$ weakly increases the marginal value of effort on average.  It is \emph{strictly} $(\mathcal{J},\mathcal{K})$-regular if the increase is strict for a positive fraction of realizations of $(a_\mathcal{J}, c_\mathcal{K})$.
\end{itemize}
\end{definition}

Weak regularity imposes monotonicity separately on each subpopulation of agents.  Strict regularity imposes the stronger requirement of strict monotonicity for a positive fraction of agents.  (In the special case of a baseline with no observed covariates, strict regularity trivially implies strict monotonicity for all agents, corresponding to Definition \ref{def:regularity}.)

The following result generalizes Theorem \ref{thm:welfare} to general baselines.

\begin{theorem}\label{thm:welfareGenBaseline}
Fix a baseline family of measured covariates $(\mathcal{J}, \mathcal{K}).$  \begin{enumerate}[(a)]
    \item For every $(\mathcal{J},\mathcal{K})$-regular attribute $j' \notin \mathcal{J}$, there exists a threshold reputation weight $\beta^* \in (0, 1]$ such that measuring $j'$ is welfare-improving if and only if $\beta > \beta^*.$ Moreover, $\beta^* < 1$ if and only if \begin{equation}
\mathbb{E}\left(MV_{\mathcal{J}\cup \{j'\},\mathcal{K}}^2\right) < \mathbb{E}\left(MV_{\mathcal{J},\mathcal{K}}^2\right)
\end{equation}
where $MV_{\mathcal{J},\mathcal{K}}$ is as defined in (\ref{def:MVexpand}).

    \item For every $(\mathcal{J},\mathcal{K})$-regular circumstance $k' \notin \mathcal{K}$, there exists a threshold reputation weight $\beta_* \in [0, 1)$ such that measuring $k'$ is welfare-improving if and only if $\beta < \beta_*.$ Moreover, $\beta_* > 0$ if and only if $k'$ is strictly $(\mathcal{J}, \mathcal{K})$-regular.
\end{enumerate}

\end{theorem}

\section{Welfare Under General Convex Costs} \label{app:WelfareCost}

Theorem \ref{thm:welfare}, our main welfare result, can be extended to non-quadratic effort cost functions under the same conditions as Proposition \ref{prop:gencorrNonQuadratic}, assuming that effort costs follow a power law.  We state and prove this result for general baselines, as in the analysis of Appendix \ref{app:GenBaseline}.

\begin{proposition}\label{prop:welfareNonQuadratic}
Suppose that $C(e) = A e^k$ for some $A > 0$ and $k > 1.$  Fix a baseline family of measured covariates $(\mathcal{J}, \mathcal{K}).$  
\begin{enumerate}[(a)]
    \item Suppose that the market measures the additional regular attribute $j' \notin \mathcal{J}.$  If either $k \geq 2$ or else all agents change their effort in the same direction, then there exists a threshold reputation weight $\beta^* \in (0, 1]$ such that the measurement is welfare-improving iff $\beta > \beta^*.$  If $j'$ is strictly regular and all agents change their effort in the same direction, then $\beta^* < 1.$

    \item Suppose that the market measures the additional regular circumstance $k' \notin \mathcal{K}.$  If either $k \leq 2$ or else all agents change their effort in the same direction, then there exists a threshold reputation weight $\beta_* \in [0, 1)$ such that the measurement is welfare-improving iff $\beta < \beta_*.$  If $k'$ is strictly regular, then $\beta_* > 0.$
\end{enumerate}
\end{proposition}

For general cost functions, fully characterizing how aggregate welfare changes with $\beta$ becomes intractable.  However, it can be shown that if the effort cost function is approximately quadratic near zero, then when $\beta$ is small, newly measured regular attributes reduce aggregate welfare and newly measured regular circumstances increase them; while for large $\beta,$ these effects reverse.

\begin{proof}[Proof of Proposition \ref{prop:welfareNonQuadratic}]  If a newly measured covariate is regular but not strictly regular, then expected effort is unchanged while the distribution of effort in each subpopulation undergoes a mean-preserving spread under the measurement.  Then since effort costs are strictly convex, aggregate welfare must at least weakly decrease no matter the value of $\beta$, corresponding to $\beta^* = 1$ for an attribute and $\beta_* = 0$ for a circumstance.  For the remainder of the proof, we assume that the newly measured attribute is strictly regular.

Let $\delta \equiv \beta/(1 - \beta).$  Note that $e^*_{\mathcal{J}, \mathcal{K}}$ depends on $\beta$ only through $\delta,$ and we will write $e^*_{\mathcal{J}, \mathcal{K}}(\delta)$ to make this dependence explicit.

We first consider the case in which the market measures a new attribute $j'.$  All notation is as in the proof of Theorem \ref{thm:welfareGenBaseline}.  Define
\[\Delta \overline{W}(\delta) \equiv \mathbb{E}[w_0(e^*_{\mathcal{J}', \mathcal{K}}(\delta)) - w_0(e^*_{\mathcal{J}, \mathcal{K}}(\delta))],\]
where
\[w_0(e) \equiv e - C(e).\]
Recall that given any family of measured covariates, equilibrium effort in a given subpopulation satisfies $e^* = (C')^{-1}(\delta \cdot MV),$ where $MV$ is the corresponding subpopulation marginal value of effort.  Thus
\[\frac{\partial e^*}{\partial \delta} = \frac{MV}{C''((C')^{-1}(\delta \cdot MV))}\]
and
\[\frac{\partial}{\partial \delta}w(e^*) = (1 - C'(e^*)) \frac{\partial e^*}{\partial \delta} = (1 - \delta \cdot MV) \cdot MV \cdot \frac{1}{C''((C')^{-1}(\delta \cdot MV))}.\]
When $C(e) = A e^k,$ we have
\[\frac{1}{C''((C')^{-1}(\delta \cdot MV))} = A' \cdot (\delta \cdot MV)^{\frac{1}{k - 1} - 1},\]
where $A' \equiv 1/((k - 1) (Ak)^{1/(k - 1)}) > 0.$  Hence
\[\frac{\partial}{\partial \delta} w(e^*) = A' \cdot \delta^{\frac{1}{k - 1} - 1} \cdot (1 - \delta \cdot MV) \cdot MV^{\frac{1}{k - 1}}.\]
Using this identity, the derivative of $\Delta \overline{W}$ may be written
\begin{align*}
\frac{\partial}{\partial \delta}\Delta \overline{W}(\delta) = A' \cdot \delta^{\frac{1}{k - 1} - 1} \cdot  \mathbb{E} & \left[ (1 - \delta \cdot MV_{\mathcal{J}', \mathcal{K}}) \cdot MV_{\mathcal{J}', \mathcal{K}}^{\frac{1}{k - 1}}\right. \\
\ & \left.- (1 - \delta \cdot MV_{\mathcal{J},\mathcal{K}}) \cdot MV_{\mathcal{J},\mathcal{K}}^{\frac{1}{k - 1}}\right].
\end{align*}
This expression crosses zero at most once for $\delta > 0$, and
\begin{align}
\lim_{\delta \rightarrow 0} \delta^{1 - \frac{1}{k - 1}} \frac{\partial}{\partial \delta} \Delta \overline{W}(\delta) = \mathbb{E}\left[MV_{\mathcal{J}', \mathcal{K}}^{\frac{1}{k - 1}} - MV_{\mathcal{J},\mathcal{K}}^{\frac{1}{k - 1}}\right]\label{eq:welfareLimLeft}
\end{align}
while
\begin{align}
\lim_{\delta \rightarrow \infty} \delta^{- \frac{1}{k - 1}} \frac{\partial}{\partial \delta} \Delta \overline{W}(\delta) = \mathbb{E}\left[MV_{\mathcal{J},\mathcal{K}}^{\frac{k}{k - 1}} - MV_{\mathcal{J}', \mathcal{K}}^{\frac{k}{k - 1}}\right]. \label{eq:welfareLimRight}
\end{align}

Strict regularity of $j'$ requires that $\mathbb{E}[MV_{\mathcal{J}', \mathcal{K}} \mid a_{\mathcal{J}}, c_{\mathcal{K}}] \leq MV_{\mathcal{J},\mathcal{K}}$ for all realizations of $(a_{\mathcal{J}}, c_{\mathcal{K}})$, with the inequality strict with positive probability.  If all agents change their effort in the same direction, $MV_{\mathcal{J}', \mathcal{K}} \leq MV_{\mathcal{J},\mathcal{K}}$ everywhere. This inequality combined with the previous implication of strict regularity implies that $MV_{\mathcal{J}', \mathcal{K}} < MV_{\mathcal{J},\mathcal{K}}$ with positive probability, in which case \eqref{eq:welfareLimLeft} is strictly negative while \eqref{eq:welfareLimRight} is strictly positive.  And in general, if $k \geq 2$ then Jensen's inequality and strict regularity imply that 
\[\mathbb{E}\left[MV_{\mathcal{J}', \mathcal{K}}^{\frac{1}{k - 1}} \mid a_{\mathcal{J}}, c_{\mathcal{K}}\right] \leq \mathbb{E}[MV_{\mathcal{J}', \mathcal{K}} \mid a_{\mathcal{J}}, c_{\mathcal{K}}]^{\frac{1}{k - 1}} \leq MV_{\mathcal{J},\mathcal{K}}^{\frac{1}{k - 1}},\]
with the final inequality strict with positive probability.  Hence
\[\mathbb{E}\left[MV_{\mathcal{J}', \mathcal{K}}^{\frac{1}{k - 1}}\right] < \mathbb{E}\left[MV_{\mathcal{J},\mathcal{K}}^{\frac{1}{k - 1}}\right].\]
Thus the derivative of $\Delta \overline{W}$ is strictly negative near zero if either $k \geq 2$ or all agents change their effort in the same direction, and in the latter case it eventually becomes positive for large $\delta.$

Next, observe that when $\delta = 0,$ $e^* = 0$ no matter the value of $MV,$ and so $\Delta \overline{W}(0) = 0.$  Then given single-crossing of the derivative of $\Delta \overline{W}(\delta),$ it must be that $\Delta \overline{W}(\delta)$ crosses zero at most once for $\delta > 0,$ and any crossing is from below.  Let $\delta^* > 0$ denote this crossing, with $\delta^* = \infty$ in the case that no crossing occurs.  Then $\Delta \overline{W}(\delta) < 0$ for $\delta < \delta^*$, while $\overline{W}(\delta) > 0$ for $\delta > \delta^*.$   Additionally, if all agents change their effort in the same direction, for large $\delta$ the derivative of $\Delta \overline{W}(\delta)$ is positive and approximately proportional to $\delta^{\frac{1}{k - 1}},$ and thus becomes unboundedly large.  It follows that eventually $\Delta \overline{W}(\delta) > 0$ for $\delta$ sufficiently large, so $\delta^* < \infty$ in this case.  Letting $\beta^* \equiv \delta^*/(1 + \delta^*)$ yields the desired reputation weight threshold.

The case of a newly measured circumstance $k'$ follows along similar lines, with two main differences.  First, strict regularity and Jensen's inequality imply that
\[\mathbb{E}\left[MV_{\mathcal{J}, \mathcal{K}'}^{\frac{k}{k - 1}} \mid a_{\mathcal{J}}, c_{\mathcal{K}}\right] \geq \mathbb{E}[MV_{\mathcal{J}, \mathcal{K}'} \mid a_{\mathcal{J}}, c_{\mathcal{K}}]^{\frac{k}{k - 1}} \geq MV_{\mathcal{J},\mathcal{K}}^{\frac{k}{k - 1}},\]
with the final inequality strict with positive probability.  Hence
\[\mathbb{E}\left[MV_{\mathcal{J},\mathcal{K}}^{\frac{k}{k - 1}} - MV_{\mathcal{J}, \mathcal{K}'}^{\frac{k}{k - 1}}\right] < 0\]
and the derivative of $\Delta \overline{W}$ is negative for large $\delta,$ whether or not all agents change their effort in the same direction.  Thus $\delta_* < \infty$ always.  Second, the parameter restriction under which Jensen's inequality implies that the derivative of $\Delta \overline{W}$ is positive near zero is $k \leq 2.$   
\end{proof}

\section{Proofs of Results from the Main Text}
\subsection{Preliminary Work: Characterization of MV}\label{app:MVChar}
Fix a family of measured covariates $(\mathcal{J}, \mathcal{K}).$  

\begin{lemma}\label{lemma:MVChar}
Suppose that Assumption \ref{ass:admissibilityGeneralBaseline} holds. Then the equilibrium marginal value of effort is
\[MV_{\mathcal{J},\mathcal{K}} = \mathbb{E}\left[\frac{\partial}{\partial Y^0} \mathbb{E}[\eta_{\mathcal{J}} \mid Y^0, a_{\mathcal{J}}, c_{\mathcal{K}}] \ \middle| \ a_{\mathcal{J}}, c_{\mathcal{K}}\right],\]
where $Y^0 \equiv \theta + \eps.$
\end{lemma}
\begin{proof}
Throughout this proof, fix a set of realizations of $(a_{\mathcal{J}}, c_{\mathcal{K}}),$ and condition all distributions on these realizations.  To economize on notation, we will suppress explicit conditioning on these covariates.  

Let $\rho_{y \mid \eta, e}(y \mid t, e)$ be the conditional density of $Y \mid \eta_{\mathcal{J}}, e$ and $\rho_{y \mid e}(y \mid e)$ be the conditional density of $Y \mid e$.   Because effort affects the outcome as an additive shift, $\rho_{y \mid \eta, e}(y \mid t, e) = \rho_{y \mid \eta, e}(y - e \mid t, 0)$ and $\rho_{y \mid e}(y \mid e) = \rho_{y \mid e}(y - e \mid 0)$ for every $(y, t, e).$  So let $\rho_{\eta \mid y, e}(t \mid y, e)$ be the conditional density of $\eta_{\mathcal{J}} \mid Y, e$, and let $\rho_{\eta}(t)$ be the conditional density of $\eta_{\mathcal{J}}$. Then by Bayes' rule,
\[\rho_{\eta \mid y, e}(t \mid y, e) = \frac{\rho_{y \mid \eta, e}(y \mid t, e) \rho_{\eta}(t)}{\rho_{y \mid e}(y \mid e)} = \frac{\rho_{y \mid \eta, e}(y - e \mid t, 0) \rho_{\eta}(t)}{\rho_{y \mid e}(y - e \mid 0)} = \rho_{\eta \mid y, e}(t \mid y - e, 0).\]
Hence 
\begin{align*}
\mathbb{E}^{e^*}[\theta \mid Y = y] = & \ \sum_{j \in \mathcal{J}} \theta_j + \mathbb{E}^{e^*}[\eta_{\mathcal{J}} \mid Y = y] \\
= & \ \sum_{j \in \mathcal{J}} \theta_j + \int t \, \rho_{\eta \mid y, e}(t \mid y, e^*) \, dt \\
= & \ \sum_{j \in \mathcal{J}} \theta_j + \int t \, \rho_{\eta \mid y, e}(t \mid y - e^*, 0) \, dt \\
= & \ \sum_{j \in \mathcal{J}} \theta_j + \mathbb{E}^{0}[\eta_{\mathcal{J}} \mid Y = y - e^*].
\end{align*}
Under the measure corresponding to $e = 0,$ the variable $Y$ is equal to $Y^0$ almost surely.  So
\[\mathbb{E}^{0}[\eta_{\mathcal{J}} \mid Y = y - e^*] = \mathbb{E}[\eta_{\mathcal{J}} \mid Y^0 = y - e^*].\]
Now, 
\begin{align*}\mathbb{E}^e\left[\mathbb{E}^{e^*}[\theta \mid Y]\right] = \ & \int dy \, \rho_{y \mid e}(y \mid e) \mathbb{E}^{e^*}[\theta \mid Y = y] \\
= \ & \int dy \, \rho_{y \mid e}(y \mid e) \left(\sum_{j \in \mathcal{J}} \theta_j + \mathbb{E}[\eta_{\mathcal{J}} \mid Y^0 = y - e^*]\right) \\
=\ & \int dy \, \rho_{y \mid e}(y - e \mid 0) \left(\sum_{j \in \mathcal{J}} \theta_j + \mathbb{E}[\eta_{\mathcal{J}} \mid Y^0 = y - e^*]\right),
\end{align*}
and so by making the variable substitution $y' = y - e$ we may write
\[\mathbb{E}^e\left[\mathbb{E}^{e^*}[\theta \mid Y]\right] = \int dy' \, \rho_{y \mid e}(y' \mid 0) \left(\sum_{j \in \mathcal{J}} \theta_j +\mathbb{E}[\eta_{\mathcal{J}} \mid Y^0 = y' - e^* + e]\right).\]
Differentiating wrt $e$ and invoking Assumption \ref{ass:admissibilityGeneralBaseline} to justify applying the dominated convergence theorem yields
\[\left.\frac{\partial}{\partial e} \mathbb{E}^e\left[\mathbb{E}^{e^*}[\theta \mid Y]\right]\right|_{e = e^*} = \int dy' \, \rho_{y \mid e}(y' \mid 0) \left.\frac{\partial}{\partial Y^0}\mathbb{E}[\eta_{\mathcal{J}} \mid Y^0]\right|_{Y^0 = y'}.\]
Recall that $Y = Y^0$ conditional on $e = 0,$ so $\rho_{y \mid e}(y' \mid 0)$ is the density of $Y^0$. The rhs of the previous expression may therefore be written
\[\left.\frac{\partial}{\partial e} \mathbb{E}^e\left[\mathbb{E}^{e^*}[\theta \mid Y]\right]\right|_{e = e^*} = \mathbb{E}\left[\frac{\partial}{\partial Y^0}\mathbb{E}[\eta_{\mathcal{J}} \mid Y^0]\right],\]
as desired.
\end{proof}

\subsection{Proofs of Theorems \ref{thm:gencorr} and \ref{thm:genBaseA}}

We prove Theorem \ref{thm:genBaseA}, from which Theorem \ref{thm:gencorr} follows immediately as a corollary. 

\subsubsection{Part (a)}
Fix a baseline family of measured covariates $(\mathcal{J}, \mathcal{K})$.  As established in Lemma \ref{lemma:MVChar}, the marginal value of effort in the baseline is
\[MV_{\mathcal{J},\mathcal{K}} = \mathbb{E}\left[\frac{\partial}{\partial Y^0} \mathbb{E}[\eta_{\mathcal{J}} \mid Y^0, a_{\mathcal{J}}, c_{\mathcal{K}}] \ \middle| \ a_{\mathcal{J}}, c_{\mathcal{K}}\right],\]
where $Y^0 \equiv \theta + \eps$ is the baseline value of the outcome after subtracting out the agent's effort.

Now suppose the market additionally observes the additional attribute $j' \notin \mathcal{J},$ and let $\mathcal{J}' \equiv \mathcal{J} \cup \{j'\}.$  Under the expanded family of measured covariates, the marginal value of effort becomes
\[MV_{\mathcal{J}', \mathcal{K}} = \mathbb{E}\left[\frac{\partial}{\partial Y^0} \mathbb{E}[\eta_{\mathcal{J}'} \mid Y^0, a_{\mathcal{J}'}, c_{\mathcal{K}}] \ \middle| \  a_{\mathcal{J}'}, c_{\mathcal{K}}\right].\]
Note that conditional on $(a_{\mathcal{J}}, c_{\mathcal{K}}),$ $MV_{\mathcal{J}', \mathcal{K}}$ is a random variable whose value is a function of the realization of $a_{j'}.$  

Because $f^{j'}$ is a one-to-one mapping, conditioning on the value of $a_{j'}$ is equivalent to conditioning on the value of $\theta_{j'} = f^{j'}(a_{j'}).$  So we may equivalently write the agent's marginal value of effort under the expanded set of covariates as
\[MV_{\mathcal{J}', \mathcal{K}} = \mathbb{E}\left[\frac{\partial}{\partial Y^0} \mathbb{E}[\eta_{\mathcal{J}'} \mid Y^0,  \theta_{j'}, a_{\mathcal{J}}, c_{\mathcal{K}}] \ \middle| \  \theta_{j'}, a_{\mathcal{J}}, c_{\mathcal{K}}\right].\]

\begin{lemma} 
 $(\eta_{\mathcal{J}}, \theta_{j'}, Y^0)$ are affiliated conditional on $(a_{\mathcal{J}}, c_{\mathcal{K}}).$ 
 \label{lemma:affiliationA2}
 \end{lemma}
 
 \begin{proof}
Fix a set of realizations of $(a_{\mathcal{J}}, c_{\mathcal{K}}),$ and conditional all distributions on their values.  To economize on notation, explicit conditioning on these covariates will be suppressed.  Let $\rho_{\eta,\theta, Y}(u, t, y)$ be the joint density of $(\eta_{\mathcal{J}}, \theta_{j'}, Y^0)$.  We will show that $\rho_{\eta,\theta, Y}$ is log-supermodular.\footnote{A vector of random variables possessing a joint density function is affiliated iff its density function is log-supermodular, that is, the logarithm of its density function is supermodular.  We make use of the following well-known facts: any product of log-supermodular functions is log-supermodular, and a twice continuously differentiable, strictly positive function $f(\textbf{x})$ is log-supermodular iff $\partial^2 \log f/\partial x_i x_j \geq 0$ for every pair of components $i \neq j$ and all $\textbf{x}.$}

Let $\rho_{\theta}(t)$ be the density of $\theta_{j'}$, $\rho_{\eta \mid \theta}(u \mid t)$ be the conditional density of $\eta_{\mathcal{J}} \mid \theta_{j'}$, and $\rho_{Y \mid \eta}(y \mid u)$ be the conditional density of $Y^0 \mid \eta_{\mathcal{J}}$.  Note that conditional on $\eta_{\mathcal{J}},$ $Y^0$ is independent of $\theta_{j'}$, and so 
\[\rho_{\eta, \theta, Y}(u, t, y) = \rho_{\theta}(t) \rho_{\eta \mid \theta}(u \mid t) \rho_{Y \mid \eta}(y \mid u).\]
It is therefore sufficient to show that $\rho_{Y \mid \eta}$ and $\rho_{\eta \mid \theta}$ are log-supermodular.  

First consider $\rho_{Y \mid \eta}$.  Define \[\mu_{(\mathcal{J}, \mathcal{K})} \equiv \sum_{j \in \mathcal{J}} \theta_j + \sum_{k \in \mathcal{K}} \eps_k + \eta_{\mathcal{J}}.\]  
Then $Y^0$ may be decomposed as
\[Y^0 = \mu_{(\mathcal{J}, \mathcal{K})} + \eta_{\mathcal{J}} + \delta_{\mathcal{K}},\]
where $\mu_{(\mathcal{J}, \mathcal{K})}$ is a constant.  Let $\rho_{\delta}(z)$ be the density of $\delta_{\mathcal{K}}$.  Then
\[\rho_{Y \mid \eta}(y \mid u) = \rho_{\delta}(y - \mu_{(\mathcal{J}, \mathcal{K})} - u).\]
Under Assumption \ref{ass:regularity}, $\rho_{\delta}$ is log-concave, meaning $\rho_{Y \mid \eta}$ is log-supermodular.

As for $\rho_{\eta \mid \theta}$, let $\rho_{\eta' \mid \theta}(w \mid t)$ be the conditional density of $\eta_{\mathcal{J}'} \mid \theta_{j'}$.  As $\eta_{\mathcal{J}} = \theta_{j'} + \eta_{\mathcal{J}'},$ it follows that 
\[\rho_{\eta \mid \theta}(u \mid t) = \rho_{\eta' \mid \theta}(u - t \mid t).\]
Hence by the chain rule,
\begin{align*}
\frac{\partial^2}{\partial u \partial t} \log \rho_{\eta \mid \theta}(u \mid t) =  \left[\frac{\partial^2}{\partial w \partial t}\log \rho_{\eta' \mid \theta}(w \mid t) - \frac{\partial^2}{\partial w^2}\log  \rho_{\eta' \mid \theta}(w \mid t)\right]_{w = u - t}. 
\end{align*}
Under Assumption \ref{ass:regularity}, $\rho_{\eta' \mid \theta}$ is log-concave and so the second term is non-negative.  Meanwhile, $\mathcal{J}$-affiliation of $j'$ implies that the first term is also non-negative.  Hence
\[\frac{\partial^2}{\partial u \partial t} \log \rho_{\eta \mid \theta}(u \mid t) \geq 0,\]
establishing the desired log-supermodularity.
\end{proof}

The marginal value of effort given observation of attribute $j'$, $MV_{\mathcal{J}', \mathcal{K}}$, can be compared to the marginal value of effort in the baseline, $MV_{\mathcal{J}', \mathcal{K}}$, as follows.  Fix a set of realizations of $(a_{\mathcal{J}}, c_{\mathcal{K}}),$ and define
\[F_\theta(t \mid y) \equiv \Pr(\theta_{j'} \leq t \mid Y^0 = y, a_{\mathcal{J}}, c_{\mathcal{K}})\]
to be the conditional distribution function of $\theta_{j'}$ given the outcome $Y^0$, and
\[\phi(y, t) \equiv \mathbb{E}[\eta_{\mathcal{J}'} \mid Y^0 = y, \theta_{j'} = t, a_{\mathcal{J}}, c_{\mathcal{K}}]\]
to be the conditional expectation of $\eta_{\mathcal{J}'}$ given $Y^0$ and $\theta_{j'}$.

By the law of total probability
\[\mathbb{E}[\eta_{\mathcal{J}} \mid Y^0 = y, a_{\mathcal{J}}, c_{\mathcal{K}}] = \int dF_\theta(t \mid y) \left(t + \phi(y, t)\right)\]
and so the change in the conditional expectation of the unobserved $\eta_{\mathcal{J}}$ as $Y^0$ moves from $y$ to $y'>y$ is
\begin{align}
& \mathbb{E}[\eta_{\mathcal{J}} \mid Y^0 = y', a_{\mathcal{J}}, c_{\mathcal{K}}] - \mathbb{E}[\eta_{\mathcal{J}} \mid Y^0 = y, a_{\mathcal{J}}, c_{\mathcal{K}}] \nonumber \\
= \ & \int d F_\theta(t \mid y') \left(t + \phi(y', t)\right) - \int d F_\theta(t \mid y) \left(t + \phi(y, t)\right) \label{exp:difference2}
\end{align}

This difference can be signed using Lemma \ref{lemma:affiliationA2}: Since $(\eta_{\mathcal{J}},\theta_{j'})$ are affiliated conditional on $Y^0$, the expression  $t + \phi(y, t)$ is nondecreasing in $t.$  And since $(\theta_{j'}, Y^0)$ are affiliated, the expectation $\mathbb{E}[\pi(\theta_{j'}) \mid Y^0, a_{\mathcal{J}}, c_{\mathcal{K}}]$ is nondecreasing in $Y^0$ for any increasing function $\pi$. Thus
\[\int dF_\theta(t \mid y') \left(t + \phi(y,t)\right)\]
is nondecreasing in $y'$, and so the expression in (\ref{exp:difference2}) can be bounded below by 
\[\int dF_\theta(t \mid y) \left(\phi(y',t) - \phi(y,t)\right).\]
It follows that
\begin{align*}
\frac{\partial}{\partial y}\mathbb{E}[\eta_{\mathcal{J}} \mid Y^0 = y, a_{\mathcal{J}}, c_{\mathcal{K}}]  \geq \int dF_\theta(t \mid y) \frac{\partial \phi}{\partial y}(y, t).
\end{align*}
The lhs is the marginal improvement in the posterior expectation of $\eta_{\mathcal{J}}$ when the realization of $Y^0$ is increased. The rhs is the expected marginal improvement in the posterior expectation of $\eta_{\mathcal{J}'}$ when it is conditioned on the manipulated realization of $Y^0$ as well as the  \emph{un-manipulated} realization of $\theta_{j'}$. 

To complete the proof, rewrite this inequality as:
\[
\frac{\partial}{\partial Y^0}\mathbb{E}[\eta_{\mathcal{J}} \mid Y^0, a_{\mathcal{J}}, c_{\mathcal{K}}] \geq \mathbb{E}\left[\frac{\partial}{\partial Y^0}\mathbb{E}[\eta_{\mathcal{J'}} \mid Y^0, \theta_{j'}, a_{\mathcal{J}}, c_{\mathcal{K}}] \ \middle| \ Y^0, a_{\mathcal{J}}, c_{\mathcal{K}}\right].\]
Taking the expectation of each side conditional on $(a_{\mathcal{J}}, c_{\mathcal{K}})$ yields
\begin{align*}
MV_{\mathcal{J},\mathcal{K}} \geq \mathbb{E}\left[\frac{\partial}{\partial Y^0}\mathbb{E}[\eta_{\mathcal{J'}} \mid Y^0, \theta_{j'}, a_{\mathcal{J}}, c_{\mathcal{K}}] \ \middle| \ a_{\mathcal{J}}, c_{\mathcal{K}}\right].
\end{align*}
By the law of iterated expectations, the rhs may be expanded as
\begin{align*}
& \mathbb{E}\left[\frac{\partial}{\partial Y^0}\mathbb{E}[\eta_{\mathcal{J'}} \mid Y^0, \theta_{j'}, a_{\mathcal{J}}, c_{\mathcal{K}}] \ \middle| \ a_{\mathcal{J}}, c_{\mathcal{K}}\right] \\
= \ & \mathbb{E}\left[\mathbb{E}\left[\frac{\partial}{\partial Y^0}\mathbb{E}[\eta_{\mathcal{J'}} \mid Y^0, \theta_{j'}, a_{\mathcal{J}}, c_{\mathcal{K}}] \ \middle| \ \theta_{j'}, a_{\mathcal{J}}, c_{\mathcal{K}}\right] \ \middle| \ a_{\mathcal{J}}, c_{\mathcal{K}}\right] \\
= \ & \mathbb{E}\left[MV_{\mathcal{J}', \mathcal{K}} \mid a_{\mathcal{J}}, c_{\mathcal{K}}\right].
\end{align*}
Therefore
\[MV_{\mathcal{J},\mathcal{K}} \geq \mathbb{E}[MV_{\mathcal{J}', \mathcal{K}} \mid a_{\mathcal{J}}, c_{\mathcal{K}}].\]
This inequality holds for every realization of $(a_{\mathcal{J}}, c_{\mathcal{K}}).$  Thus the marginal value of effort in every $(\mathcal{J}, \mathcal{K})$-subpopulation is weakly larger than the expected marginal value once attribute $j'$ is additionally measured, as desired.

\subsubsection{Part (b)}

Suppose that the market observes the additional circumstance $k' \notin \mathcal{K}.$  Let $\mathcal{K}' \equiv \mathcal{K} \cup \{k'\}.$  Under the expanded family of measured covariates, the marginal value of effort becomes
\[MV_{\mathcal{J}, \mathcal{K}'} = \mathbb{E}\left[\frac{\partial}{\partial Y^0} \mathbb{E}[\eta_{\mathcal{J}'} \mid Y^0, a_{\mathcal{J}}, c_{\mathcal{K}'}] \ \middle| \  a_{\mathcal{J}}, c_{\mathcal{K}'}\right].\]
Because $g^{k'}$ is a one-to-one-mapping, conditioning on $c_{k'}$ is equivalent to conditioning on $\eps_{k'} = g^{k'}(c_{k'})$.  We may therefore equivalently write the agent's marginal value of effort under the expanded set of covariates as
\[MV_{\mathcal{J}, \mathcal{K}'} = \mathbb{E}\left[\frac{\partial}{\partial Y^0} \mathbb{E}[\eta_{\mathcal{J}'} \mid Y^0, \eps_{k'}, a_{\mathcal{J}}, c_{\mathcal{K}}] \ \middle| \  \eps_{k'}, a_{\mathcal{J}}, c_{\mathcal{K}}\right].\]

\begin{lemma} \label{lemma:affiliationC2}
$(\eps_{k'}, \delta_{\mathcal{K}}, Y^0)$ are affiliated conditional on $(a_{\mathcal{J}}, c_{\mathcal{K}}).$
\end{lemma}
\begin{proof}
This is established along very similar lines to the proof of Lemma \ref{lemma:affiliationA2}.  Fix realizations of $(a_{\mathcal{J}}, c_{\mathcal{K}}),$ condition all distributions on their values, and suppress explicit conditioning.  Let $\rho_{\eta}(u)$ be the density of $\eta_{\mathcal{J}}$ and $\rho_{\delta' \mid \eps}(x \mid z)$ be the conditional density of $\delta_{-\mathcal{K}'} \mid \eps_{k'}$.  The conditions required for the steps of the proof of Lemma \ref{lemma:affiliationA2} to go through are that $\rho_{\eta}(u)$ is log-concave, $\rho_{\delta' \mid \eps}(x \mid z)$ is log-concave in $x$ for all $z,$ and $(\delta_{\mathcal{K}'}, \eps_{k'})$ are affiliated.  The first two properties follow from Assumption \ref{ass:admissibilityGeneralBaseline}, while the final property holds by $\mathcal{K}$-affiliation of circumstance $k'.$
\end{proof}

We compare $MV_{\mathcal{J}, \mathcal{K}'}$ with $MV_{\mathcal{J},\mathcal{K}}$ in a manner very similar to the case of an additional attribute.  Fix realizations of $(a_{\mathcal{J}}, c_{\mathcal{K}}),$ and  define
\[F_\eps(z \mid y) \equiv \Pr(\eps_{k'} \leq z \mid Y^0 = y, a_{\mathcal{J}}, c_{\mathcal{K}})\]
to be the conditional CDF of $\eps_{k'}$ given the outcome $Y^0$.  Decompose $Y^0$ as
\[Y^0 = \mu_{(\mathcal{J}, \mathcal{K})} + \eta_{\mathcal{J}} + \delta_{\mathcal{K}}.\]
Taking expectations of each side conditional on $(Y^0, \eps_{k'}, a_{\mathcal{J}}, c_{\mathcal{K}})$ yields
\[Y^0 = \mu_{(\mathcal{J}, \mathcal{K})} + \mathbb{E}[\eta_{\mathcal{J}} \mid Y^0, \eps_{k'}, a_{\mathcal{J}}, c_{\mathcal{K}}] + \mathbb{E}[\delta_{\mathcal{K}} \mid Y^0, \eps_{k'}, a_{\mathcal{J}}, c_{\mathcal{K}}].\]
Hence
\begin{align}
& \int dF_\eps(z \mid y) \mathbb{E}[\eta_{\mathcal{J}} \mid Y^0 = y, \eps_{k'} = z, a_{\mathcal{J}}, c_{\mathcal{K}}] \nonumber \\
= \ & y - \mu_{(\mathcal{J}, \mathcal{K})} - \int dF_\eps(z \mid y) \mathbb{E}[\delta_{\mathcal{K}} \mid Y^0 = y, \eps_{k'} = z, a_{\mathcal{J}}, c_{\mathcal{K}}]. \label{eq:expC2}
\end{align}
Lemma \ref{lemma:affiliationC2} directly implies that
\[\int dF_\eps(z \mid y) \mathbb{E}[\delta_{\mathcal{K}} \mid Y^0 = y, \eps_{k'} = z, a_{\mathcal{J}}, c_{\mathcal{K}}]\]
is weakly increasing in $y$, so (\ref{eq:expC2}) is weakly decreasing in $y$.

Following the same logic as in the attributes case, monotonicity of (\ref{eq:expC2}) implies that
\[
\frac{\partial}{\partial Y^0}\mathbb{E}[\eta_{\mathcal{J}} \mid Y^0, a_{\mathcal{J}}, c_{\mathcal{K}}] \leq \mathbb{E}\left[\frac{\partial}{\partial Y^0}\mathbb{E}[\eta_{\mathcal{J}} \mid Y^0, \eps_{k'}, a_{\mathcal{J}}, c_{\mathcal{K}}] \ \middle| \ Y^0, a_{\mathcal{J}}, c_{\mathcal{K}}\right],\]
and it follows that
\[MV_{\mathcal{J},\mathcal{K}} \leq \mathbb{E}[MV_{\mathcal{J}, \mathcal{K}'} \mid a_{\mathcal{J}}, c_{\mathcal{K}}].\]
Thus the marginal value of effort in each subpopulation in the baseline is weakly lower than the expected marginal value of effort when the circumstance $k'$ is additionally measured.

\subsection{Proofs of Theorems \ref{thm:mean-shift} and \ref{thm:genBaseMS}} \label{proof:mean-shift}

We prove Theorem \ref{thm:genBaseMS}, from which Theorem \ref{thm:mean-shift} follows immediately as a corollary. 
\subsubsection{Part (a)}
Fix a baseline family of measured covariates $(\mathcal{J}, \mathcal{K})$.  As established in Lemma \ref{lemma:MVChar}, the marginal value of effort is
\[MV(\mathcal{J}, \mathcal{K}) = \mathbb{E}\left[\frac{\partial}{\partial Y^0} \mathbb{E}[\eta_{\mathcal{J}} \mid Y^0, a_{\mathcal{J}}, c_{\mathcal{K}}] \ \middle| \ a_{\mathcal{J}}, c_{\mathcal{K}}\right],\]
where $Y^0 \equiv \theta + \eps$ is the baseline value of the outcome after subtracting out the agent's effort.

Now suppose the market additional observes the additional attribute $j' \notin \mathcal{J},$ and let $\mathcal{J}' \equiv \mathcal{J} \cup \{j'\}.$  Under the expanded family of measured covariates, the marginal value of effort becomes
\[MV_{\mathcal{J}', \mathcal{K}} = \mathbb{E}\left[\frac{\partial}{\partial Y^0} \mathbb{E}[\eta_{\mathcal{J}'} \mid Y^0, a_{\mathcal{J}'}, c_{\mathcal{K}}] \ \middle| \ a_{\mathcal{J}'}, c_{\mathcal{K}}\right],\]
where, conditional on $(a_{\mathcal{J}}, c_{\mathcal{K}})$, $MV_{\mathcal{J}', \mathcal{K}}$ is a random variable whose value is a function of the realization of $a_{j'}.$  

The outcome $Y^0$ may be decomposed as 
\begin{equation} \label{eq:Y0}
Y^0 = \mu_{(\mathcal{J}, \mathcal{K})} + \eta_{\mathcal{J}} + \delta_{\mathcal{K}},
\end{equation}
 where \[\mu_{(\mathcal{J}, \mathcal{K})} \equiv \sum_{j \in \mathcal{J}} \theta_j + \sum_{k \in \mathcal{K}} \eps_k\] is constant conditional on $(a_{\mathcal{J}'}, c_{\mathcal{K}})$.  The residual type component $\eta_{\mathcal{J}}$ may be further decomposed as
 \[\eta_{\mathcal{J}} = \bar{\eta}_{\mathcal{J} \mid j'} + \Delta \eta_{\mathcal{J}'},\]
 where
 \[\bar{\eta}_{\mathcal{J} \mid j'} \equiv \mathbb{E}[\eta_{\mathcal{J}} \mid a_{\mathcal{J}'}], \quad \Delta \eta_{\mathcal{J}'} \equiv \theta - \mathbb{E}[\theta \mid a_{\mathcal{J}'}].\]
We may therefore rewrite (\ref{eq:Y0}) as
\[Y^0 = \mu_{(\mathcal{J}, \mathcal{K})} + \bar{\eta}_{\mathcal{J} \mid j'} + \Delta \eta_{\mathcal{J}'} + \delta_{\mathcal{K}}.\]

Now, note that
\[\eta_{\mathcal{J}'} - \mathbb{E}[\eta_{\mathcal{J}'} \mid a_{\mathcal{J}'}] = \sum_{j \in \mathcal{J}'} \theta_{j'} + \eta_{\mathcal{J}'} - \mathbb{E}\left[\sum_{j \in \mathcal{J}'} \theta_{j'} + \eta_{\mathcal{J}'} \mid a_{\mathcal{J}'}\right] = \Delta \eta_{\mathcal{J}'}.\]
Hence $\mathcal{J}$-Strong Homoskedasticity of attribute $j'$ is equivalent to the assumption that $\Delta \eta_{\mathcal{J}'}$ is independent of $a_{j'}$ conditional on $(a_{\mathcal{J}}, c_{\mathcal{K}}).$  Therefore under $\mathcal{J}$-Strong Homoskedasticity, $Y^0$ depends on $a_{j'}$ only through $\bar{\eta}_{\mathcal{J} \mid j'}.$  It follows that under $\mathcal{J}$-Strong Homoskedasticity, $\mathbb{E}[\Delta \eta_{\mathcal{J}'} \mid Y^0, a_{\mathcal{J}'}, c_{\mathcal{K}}] = \mathbb{E}[\Delta \eta_{\mathcal{J}'} \mid Y^0, \bar{\eta}_{\mathcal{J} \mid j'}, a_{\mathcal{J}}, c_{\mathcal{K}}]$, and the latter expectation depends on $a_{j'}$ only through $\bar{\eta}_{\mathcal{J} \mid j'}$.

Using this fact, we may write
\[
\mathbb{E}[\eta_{\mathcal{J}'} \mid Y^0, a_{\mathcal{J}'}, c_{\mathcal{K}}] = \bar{\eta}_{\mathcal{J} \mid j'} + \mathbb{E}[\Delta \eta_{\mathcal{J}'} \mid Y^0, \bar{\eta}_{\mathcal{J} \mid j'}, a_{\mathcal{J}}, c_{\mathcal{K}}]\]
and 
\[MV_{\mathcal{J}', \mathcal{K}} = \mathbb{E}\left[\frac{\partial}{\partial Y^0} \mathbb{E}[\Delta \eta_{\mathcal{J}'} \mid Y^0, \bar{\eta}_{\mathcal{J} \mid j'}, a_{\mathcal{J}}, c_{\mathcal{K}}] \ \middle| \  \bar{\eta}_{\mathcal{J} \mid j'}, a_{\mathcal{J}}, c_{\mathcal{K}}\right].\]

The theorem holds if can we show that the conditional expectation of  $\Delta \eta_{\mathcal{J}'}$ is less responsive to the realization of the outcome $Y$ than the conditional expectation of the original residual  $\eta_{\mathcal{J}}$. Note that $\eta_{\mathcal{J}}$ is the sum of the (conditionally) independent variables $\bar{\eta}_{\mathcal{J} \mid j'}$ and  $\Delta \eta_{\mathcal{J}'}$, so uncertainty about $\Delta \eta_{\mathcal{J}'}$ is mechanically lower than uncertainty about $\eta_{\mathcal{J}}$. But this does not directly translate into a statement that the posterior expectation of $\Delta \eta_{\mathcal{J}'}$ is less sensitive to the realization of $Y$. In general, we are not even guaranteed that higher realizations of $Y$ lead to higher inferences about $\Delta \theta_{\mathcal{J}}$ once we have conditioned on the realization of $\bar{\eta}_{\mathcal{J} \mid j'}$.\footnote{Recall that our admissibility assumptions are imposed on the original type component $\theta_{j'}$, and not on the constructed $\bar{\eta}_{\mathcal{J} \mid j'}$.} We next prove a  key technical lemma, which will imply an analogue of admissibility for our transformed environment. 

\begin{lemma} $(\eta_{\mathcal{J}}, \bar{\eta}_{\mathcal{J} \mid j'}, Y^0)$ are affiliated conditional on $(a_{\mathcal{J}}, c_{\mathcal{K}}).$ \label{lemm:Affiliation}
\end{lemma}

\begin{proof}
Fix a set of realizations of $(a_{\mathcal{J}}, c_{\mathcal{K}}),$ and condition all distributions on these values.  To economize on notation, we suppress explicit conditioning on these covariates throughout this proof.  Let $\tilde{\rho}_{\eta, \theta, Y}(u, t, y)$ be the conditional joint density of $(\eta_{\mathcal{J}}, \bar{\eta}_{\mathcal{J} \mid j'}, Y^0)$.  We will show that $\tilde{\rho}_{\eta, \theta, Y}$ is log-supermodular.

Use $\tilde{\rho}_{\theta}(t)$ to denote the density of $\bar{\eta}_{\mathcal{J} \mid j'}$, $\tilde{\rho}_{\eta \mid \theta}(u \mid t)$ to denote the conditional density of $\eta_{\mathcal{J}}\mid \bar{\eta}_{\mathcal{J} \mid j'}$, and $\tilde{\rho}_{Y \mid \eta}(y \mid u)$ to denote the conditional density of $Y^0 \mid \eta_{\mathcal{J}}$.  Note that $Y^0$ is independent of $\bar{\eta}_{\mathcal{J} \mid j'}$ conditional on $\eta_{\mathcal{J}}$.  So $\tilde{\rho}_{\eta, \theta, Y}$ may be decomposed as 
\[\tilde{\rho}_{\eta, \theta, Y}(u, t, y) = \tilde{\rho}_{\theta}(t) \tilde{\rho}_{\eta \mid \theta}(u \mid t) \tilde{\rho}_{Y \mid \eta}(y \mid u).\]
It is therefore sufficient to show that $\tilde{\rho}_{Y \mid \eta}$ and $\tilde{\rho}_{\eta \mid \theta}$ are log-supermodular.  

First consider $\tilde{\rho}_{Y \mid \eta}$.  Decompose $Y^0$ as
\[Y^0 = \mu_{(\mathcal{J}, \mathcal{K})} + \eta_{\mathcal{J}} + \delta_{\mathcal{K}}.\]
Let $\rho_{\delta}(z)$ be the density of $\delta_{\mathcal{K}}$. Then
\[\tilde{\rho}_{Y \mid \eta}(y \mid u) = \rho_{\delta}(y - \mu_{(\mathcal{J}, \mathcal{K})} - u).\]
Under Assumption \ref{ass:admissibilityGeneralBaseline}, $\rho_{\delta}$ is log-concave, meaning $\tilde{\rho}_{Y \mid \eta}$ is log-supermodular.

As for $\tilde{\rho}_{\eta \mid \theta}$, let $\tilde{\rho}_{\eta'}(w)$ be the density of $\Delta \eta_{\mathcal{J}'}$.  Decompose $\eta_{\mathcal{J}}$ as \[\eta_{\mathcal{J}} = \bar{\eta}_{\mathcal{J} \mid j'} + \Delta \eta_{\mathcal{J}'},\] and recall that if $j'$ is $\mathcal{J}$-Strongly Homoskedastic, then $\Delta \eta_{\mathcal{J}'}$ is independent of $a_{j'}$ and hence $\bar{\eta}_{\mathcal{J} \mid j'}$.  It follows that
\[\tilde{\rho}_{\eta \mid \theta}(u \mid t) = \tilde{\rho}_{\eta'}(u - t),\]
and hence
\begin{align*}
\frac{\partial^2}{\partial u \partial t} \log \tilde{\rho}_{\eta \mid \theta}(u \mid t) =  - \left.\frac{\partial^2}{\partial w^2}\log  \tilde{\rho}_{\eta'}(w)\right|_{w = u - t} = -\frac{\partial^2}{\partial u^2}\log  \tilde{\rho}_{\eta \mid \theta}(u \mid t). 
\end{align*}

Now, let $\rho_{\eta \mid a}(u \mid \alpha)$ denote the conditional density of $\eta_{\mathcal{J}} \mid a_{j'}$.  Define 
\[\zeta(\alpha) \equiv f^{j'}(\alpha) + \mathbb{E}[\eta_{\mathcal{J}'} \mid a_{j'} = \alpha],\]
so that
\[\eta_{\mathcal{J}} = \zeta(a_{j'}) + \Delta \eta_{\mathcal{J}'}.\]
Strong Homoskedasticity implies that
\[\rho_{\eta \mid a}(u \mid \alpha) = \tilde{\rho}_{\eta'}(u - \zeta(\alpha)) = \tilde{\rho}_{\eta \mid \theta}(u \mid \zeta(\alpha)).\]
Let $\tilde{\Theta} \equiv \{t \ : \ \zeta(\alpha) = t \text{ for some } \alpha \in A_{j'}\}$ denote the support of $\bar{\eta}_{\mathcal{J} \mid j'}.$  Fix any $t \in \tilde{\Theta}.$  Then for all $u$ and every $\alpha \in A_{j'}$ such that $\zeta(\alpha) = t,$
\[\frac{\partial^2}{\partial u^2}\log  \tilde{\rho}_{\eta \mid \theta}(u \mid t) = \frac{\partial^2}{\partial u^2}\log  \rho_{\eta \mid a}(u \mid \alpha).\]
Let $\rho_{\eta' \mid a}$ denote the conditional density of $\eta_{\mathcal{J}'} \mid a_{j'}$.  Then $\rho_{\eta \mid a}(u \mid \alpha) = \rho_{\eta' \mid a}(u - f^{j'}(\alpha) \mid \alpha),$ so that
\[\frac{\partial^2}{\partial u^2}\log  \tilde{\rho}_{\eta \mid \theta}(u \mid t) = \left.\frac{\partial^2}{\partial w^2}\log  \rho_{\eta' \mid a}(w \mid \alpha)\right|_{w = u - f^{j'}(\alpha)}.\]
Under Assumption \ref{ass:admissibilityGeneralBaseline}, $\rho_{\eta' \mid a}$ is log-concave and this final derivative is non-positive, meaning
\[\frac{\partial^2}{\partial u \partial t} \log \tilde{\rho}_{\eta \mid \theta}(u \mid t) = -\frac{\partial^2}{\partial u^2}\log  \tilde{\rho}_{\eta \mid \theta}(u \mid t) \geq 0\]
for every $u$ and $t \in \tilde{\Theta}.$  Hence $\tilde{\rho}_{\eta \mid \theta}$ is log-supermodular, as desired. \end{proof}

Following arguments identical to those used 
for the proof of Theorem \ref{thm:genBaseA} (with $\Delta \eta_{\mathcal{J}'}$ and $\bar{\eta}_{\mathcal{J} \mid j'}$ playing the roles of $\eta_{\mathcal{J}'}$ and $\theta_{j'}$), Lemma \ref{lemm:Affiliation} implies
\[MV_{\mathcal{J},\mathcal{K}} \geq \mathbb{E}[MV_{\mathcal{J}', \mathcal{K}} \mid a_{\mathcal{J}}, c_{\mathcal{K}}].\]
Thus the marginal value of effort in each subpopulation in the baseline setting is weakly higher than the expected marginal value once attribute $j'$ is additionally measured, for any realizations of $(a_{\mathcal{J}}, c_{\mathcal{K}})$.

To complete the proof, we must establish that monotonicity holds uniformly across realizations of $a_{j'}$, and not just on average.  This follows immediately from the fact that $MV_{\mathcal{J}', \mathcal{K}}$ is independent of $a_{j'}$ conditional on $(a_{\mathcal{J}}, c_{\mathcal{K}})$.   To see this, decompose $Y$ as
\[Y = e + \mu_{(\mathcal{J}, \mathcal{K})} + \bar{\eta}_{\mathcal{J} \mid j'} + \Delta \eta_{\mathcal{J}'} + \delta_{\mathcal{K}}.\]
Strong Homoskedasticity implies that $\Delta \eta_{\mathcal{J}'}$ is independent of $a_{j'}$ conditional on $(a_{\mathcal{J}}, c_{\mathcal{K}})$.  Hence $a_{j'}$ enters the market's inference problem only via a known additive shift $\bar{\eta}_{\mathcal{J} \mid j'} $ to the agent's type distribution, and therefore its value does not impact incentives for effort.  So these incentives must be independent of $a_{j'},$ as claimed.

\subsubsection{Part (b)}
Suppose the market observes the additional circumstance $k' \notin \mathcal{K}.$ Let $\mathcal{K}' \equiv \mathcal{K} \cup \{k'\}.$ Under the expanded family of measured covariates, the marginal value of effort is
\[MV_{\mathcal{J}, \mathcal{K}'} = \mathbb{E}\left[\frac{\partial}{\partial Y^0} \mathbb{E}[\eta_{\mathcal{J}} \mid Y^0, a_{\mathcal{J}}, c_{\mathcal{K}'}] \ \middle| \   a_{\mathcal{J}}, c_{\mathcal{K}'}\right].\]
Define \[\bar{\delta}_{\mathcal{K}, k'} \equiv  \mathbb{E}[\delta_{\mathcal{K}} \mid c_{\mathcal{K}'}], \quad \Delta \delta_{\mathcal{K}'} \equiv \eps - \mathbb{E}[\eps \mid c_{\mathcal{K}'}].\]
Then $Y^0$ may be decomposed as
\[Y^0 = \mu_{(\mathcal{J}, \mathcal{K})} + \eta_{\mathcal{J}} + \bar{\delta}_{\mathcal{K}, k'} + \Delta \delta_{\mathcal{K}'}.\]
Under Strong Homoskedasticity, $\Delta \delta_{\mathcal{K}'}$ is independent of $c_{k'}$ conditional on $(a_{\mathcal{J}}, c_{\mathcal{K}}),$ and so the distribution of $Y^0$ depends on $c_{k'}$ only through $\bar{\delta}_{\mathcal{K}, k'}.$  Thus
\[\mathbb{E}[\eta_{\mathcal{J}} \mid Y^0, a_{\mathcal{J}}, c_{\mathcal{K}'}] = \mathbb{E}[\eta_{\mathcal{J}} \mid Y^0, \bar{\delta}_{\mathcal{K}, k'}, a_{\mathcal{J}}, c_{\mathcal{K}}],\]
and the latter random variable depends on $c_{k'}$ only through $\bar{\delta}_{\mathcal{K}, k'}.$  Therefore, in a manner analogous to the attribute case, the marginal value of effort after measuring $j'$ depends on $c_{k'}$ only through $\bar{\delta}_{\mathcal{K}, k'}$ and may be written
\[MV_{\mathcal{J}, \mathcal{K}'} = \mathbb{E}\left[\frac{\partial}{\partial Y^0} \mathbb{E}[\eta_{\mathcal{J}} \mid Y^0, \bar{\delta}_{\mathcal{K}, k'}, a_{\mathcal{J}}, c_{\mathcal{K}}] \ \middle| \  \bar{\delta}_{\mathcal{K}, k'}, a_{\mathcal{J}}, c_{\mathcal{K}}\right].\]

\begin{lemma} \label{lemm:pfAffiliationC1} $(\bar{\delta}_{\mathcal{K}, k'}, \delta_{\mathcal{K}}, Y^0)$ are affiliated conditional on $(a_{\mathcal{J}}, c_{\mathcal{K}})$.
\end{lemma}

\begin{proof}
This proof follows along very similar lines to the proof of Lemma \ref{lemm:Affiliation}.  Fix realizations of $(a_{\mathcal{J}}, c_{\mathcal{K}}),$ condition all distributions on their values, and suppress explicit conditioning on these covariates.  Let $\rho_{\eta}(u)$ be the conditional density of $\eta_{\mathcal{J}}$ and $\rho_{\delta' \mid c}(x \mid z)$ be the conditional density of $\delta_{\mathcal{K}'} \mid c_{k'}$.  The conditions required for the steps of the proof of Lemma \ref{lemm:Affiliation} to go through are that $\rho_{\eta}(u)$ is log-concave, $\rho_{\delta' \mid c}(x \mid z)$ is log-concave in $x$ for all $z,$ and $\Delta \delta_{\mathcal{K}'}$ is independent of $\bar{\delta}_{\mathcal{K} \mid k'}$.  The first two properties are ensured by Assumption \ref{ass:admissibilityGeneralBaseline}, while the final property holds under Strong Homoskedasticity.
\end{proof}

We compare $MV_{\mathcal{J}, \mathcal{K}'}$ and $MV_{\mathcal{J},\mathcal{K}}$ in a manner very similar to the attribute case.  Fix realizations of $(a_{\mathcal{J}}, c_{\mathcal{K}}).$  Define
\[\widetilde{F}_\eps(z \mid y) \equiv \Pr(\bar{\delta}_{\mathcal{K}, k'} \leq z \mid Y^0 = y, a_{\mathcal{J}}, c_{\mathcal{K}})\]
to be the distribution function of $\bar{\delta}_{\mathcal{K}, k'}$ conditional on the outcome $Y^0$.

Decompose $Y^0$ as
\[Y^0 = \mu_{(\mathcal{J}, \mathcal{K})} + \eta_{\mathcal{J}} + \delta_{\mathcal{K}}.\]
Taking expectations of each side conditional on $(Y^0, \bar{\delta}_{\mathcal{K}, k'}, a_{\mathcal{J}}, c_{\mathcal{K}})$ yields
\[Y^0 = \mu_{(\mathcal{J}, \mathcal{K})} + \mathbb{E}[\eta_{\mathcal{J}} \mid Y^0, \bar{\delta}_{\mathcal{K}, k'}, a_{\mathcal{J}}, c_{\mathcal{K}}] + \mathbb{E}[\delta_{\mathcal{K}} \mid Y^0, \bar{\delta}_{\mathcal{K}, k'}, a_{\mathcal{J}}, c_{\mathcal{K}}].\]
Hence
\begin{align}
& \int d\widetilde{F}_\eps(z \mid y') \mathbb{E}[\eta_{\mathcal{J}} \mid Y^0 = y, \bar{\delta}_{\mathcal{K}, k'} = z, a_{\mathcal{J}}, c_{\mathcal{K}}] \nonumber \\
= \ & y - \mu_{(\mathcal{J}, \mathcal{K})} - \int d\widetilde{F}_\eps(z \mid y') \mathbb{E}[\delta_{\mathcal{K}} \mid Y^0 = y, \bar{\delta}_{\mathcal{K}, k'} = z, a_{\mathcal{J}}, c_{\mathcal{K}}]. \label{eq:expC}
\end{align}
Lemma \ref{lemm:pfAffiliationC1} directly implies that 
\[\int d\widetilde{F}_\eps(z \mid y) \mathbb{E}[\delta_{\mathcal{K}} \mid Y^0 = y, \bar{\delta}_{\mathcal{K}, k'} = z, a_{\mathcal{J}}, c_{\mathcal{K}}]\]
is nondecreasing in $y$, so (\ref{eq:expC}) is nonincreasing in $y$.

Following the same logic as in the attributes case, monotonicity of (\ref{eq:expC}) implies that
\[
\frac{\partial}{\partial Y^0}\mathbb{E}[\eta_{\mathcal{J}} \mid Y^0, a_{\mathcal{J}}, c_{\mathcal{K}}] \leq \mathbb{E}\left[\frac{\partial}{\partial Y^0}\mathbb{E}[\eta_{\mathcal{J}} \mid Y^0, \bar{\delta}_{\mathcal{K}, k'}, a_{\mathcal{J}}, c_{\mathcal{K}}] \ \middle| \ Y^0, a_{\mathcal{J}}, c_{\mathcal{K}}\right],\]
and it follows that
\[MV_{\mathcal{J},\mathcal{K}} \leq \mathbb{E}[MV_{\mathcal{J}, \mathcal{K}'} \mid a_{\mathcal{J}}, c_{\mathcal{K}}].\]
Thus the marginal value of effort in each subpopulation in the baseline is weakly lower than the expected marginal value of effort when the circumstance $k'$ is additionally measured.

The final step in the proof is to establish that monotonicity holds uniformly across realizations of the additional circumstance, and not just on average.  This follows from nearly identical work to the argument for the attributes case.

\subsection{Proof of Theorem \ref{thm:welfare}}
Define $\delta \equiv \beta/(1 - \beta).$  Then equilibrium effort under any measured covariates can be written as $e^*_{\mathcal{J}, \mathcal{K}} = \delta \cdot MV_{\mathcal{J}, \mathcal{K}}.$  We will characterize the aggregate welfare change from measuring a new covariate as a function of $\delta$, which can be mapped onto an equivalent characterization in terms of $\beta.$

Consider first the case in which the market observes the new attribute $j'.$  Let $\mathcal{J}' \equiv \mathcal{J} \cup \{j'\}.$  The aggregate change in welfare from measuring $j',$ as a function of $\delta,$ is
\[\Delta \overline{W}(\delta) = \mathbb{E}\left[w_0(\delta \cdot MV_{\mathcal{J}', \mathcal{K}}) - w_0(\delta \cdot MV_{\mathcal{J},\mathcal{K}})\right],\]
where
\[w_0(e) \equiv e - \frac{1}{2} e^2.\]
The aggregate change in welfare is a quadratic function of $\delta,$ whose first derivative is
\[
\frac{d}{d\delta} \Delta \overline{W}(\delta) =  \mathbb{E} \left[ MV_{\mathcal{J}', \mathcal{K}}  - MV_{\mathcal{J},\mathcal{K}}\right] - \delta \cdot \mathbb{E}\left[ MV_{\mathcal{J}', \mathcal{K}}^2  - MV_{\mathcal{J},\mathcal{K}}^2 \right].\]
Observe that  $\overline{W}(0) = 0$ and 
\[\left.\frac{d}{d\delta} \Delta \overline{W}(\delta) \right|_{\delta = 0}  = \mathbb{E}\left[ MV_{\mathcal{J}', \mathcal{K}} - MV_{\mathcal{J},\mathcal{K}}\right] \leq 0\]
with the inequality implied by regularity of $j'$. So $\Delta \overline{W}(\delta)$ is a quadratic function that vanishes at $\delta = 0$ and is nonincreasing there.  Suppose first that $j'$ is strictly regular, so that $\Delta \overline{W}$ is strictly decreasing at 0.  Then its shape must then satisfy either of:
\begin{itemize}
    \item $\Delta \overline{W}(\delta)$ is strictly convex and intersects zero exactly once for $\delta > 0,$
    \item $\Delta \overline{W}(\delta)$ is weakly concave and does not intersect zero for any $\delta > 0.$
\end{itemize}
Define $\delta^* \equiv \inf\{\delta > 0 \ : \ \Delta \overline{W}(\delta) \geq 0\}$.  This threshold lies in $(0, \infty),$ and has the property that $\Delta \overline{W}(\delta) > 0$ for $\delta > \delta^*,$ and $\Delta \overline{W}(\delta) \leq 0$ for $\delta \in (0, \delta^*]$.  Further, $\delta^* < \infty$ if and only if $\Delta \overline{W}$ is strictly convex.  Convexity is determined by the sign of
\[\frac{d^2}{d\delta^2}\Delta \overline{W}(\delta) = \mathbb{E}\left[MV_{\mathcal{J},\mathcal{K}}^2 - MV_{\mathcal{J}', \mathcal{K}}^2\right].\]
Hence $\delta^* < \infty$ if and only if $\mathbb{E}\left[MV_{\mathcal{J},\mathcal{K}}^2\right] > \mathbb{E}\left[MV_{\mathcal{J}', \mathcal{K}}^2\right].$  Letting $\beta^* \equiv \delta^*/(1 + \delta^*)$ therefore yields a reputation weight threshold with the desired properties.

On the other hand, suppose that $j'$ is weakly but not strictly regular.  Write
\begin{align*}
& \mathbb{E}[MV_{\mathcal{J}', \mathcal{K}}^2 \mid a_{\mathcal{J}}, c_{\mathcal{K}}] \\
= \ & \mathbb{E}[(MV_{\mathcal{J}', \mathcal{K}} - \mathbb{E}[MV_{\mathcal{J}', \mathcal{K}} \mid a_{\mathcal{J}}, c_{\mathcal{K}}])^2\mid a_{\mathcal{J}}, c_{\mathcal{K}}] + \mathbb{E}[MV_{\mathcal{J}', \mathcal{K}} \mid a_{\mathcal{J}}, c_{\mathcal{K}}]^2 \\
= \ & \text{Var}\left(MV_{\mathcal{J}', \mathcal{K}} \mid a_{\mathcal{J}}, c_{\mathcal{K}}\right) + \mathbb{E}[MV_{\mathcal{J}', \mathcal{K}} \mid a_{\mathcal{J}}, c_{\mathcal{K}}]^2.
\end{align*}
Hence
\begin{align*}
& \mathbb{E}[ MV_{\mathcal{J}', \mathcal{K}}^2 \mid a_{\mathcal{J}}, c_{\mathcal{K}}] - MV_{\mathcal{J},\mathcal{K}}^2 \\
= \ & \mathbb{E}[MV_{\mathcal{J}', \mathcal{K}} \mid a_{\mathcal{J}}, c_{\mathcal{K}}]^2 - MV_{\mathcal{J},\mathcal{K}}^2 + \text{Var}\left(MV_{\mathcal{J}', \mathcal{K}} \mid a_{\mathcal{J}}, c_{\mathcal{K}}\right).
\end{align*}
Weak but not strict regularity implies that $MV_{\mathcal{J},\mathcal{K}} = \mathbb{E}[MV_{\mathcal{J}', \mathcal{K}} \mid a_{\mathcal{J}}, c_{\mathcal{K}}]$ with probability 1, so that
\[\mathbb{E}[ MV_{\mathcal{J}', \mathcal{K}}^2 \mid a_{\mathcal{J}}, c_{\mathcal{K}}] - MV_{\mathcal{J},\mathcal{K}}^2 = \text{Var}\left(MV_{\mathcal{J}', \mathcal{K}} \mid a_{\mathcal{J}}, c_{\mathcal{K}}\right) \geq 0\]
with probability 1.  Therefore $\Delta \overline{W}$ is at least weakly concave.  Weak but not strict regularity additionally implies that $\Delta \overline{W}$ has zero slope at $\delta = 0.$ Hence in this case $\Delta \overline{W}(\delta) \leq 0$ for all $\delta > 0$, in which case $\delta^* = \infty$ and $\beta^* = \infty.$  
We may summarize the work for the strictly and weakly regular cases by the result that $\beta^* < \infty$ if and only if $\mathbb{E}\left[MV_{\mathcal{J},\mathcal{K}}^2\right] > \mathbb{E}\left[MV_{\mathcal{J}', \mathcal{K}}^2\right]$.  Indeed, in the strictly regular case, this equivalence was directly established, while in the weakly regular case, it is true both that $\beta^* = \infty$ and that the latter inequality cannot hold.

Now suppose the market observes the new circumstance $k'.$  Calculations very similar to those for the attribute case show that $\Delta \overline{W}$ is a quadratic function of $\delta$ which vanishes at $\delta = 0$ and is weakly increasing there, and strictly increasing if $k'$ is strictly regular.  It is concave iff $\mathbb{E}\left[MV_{\mathcal{J},\mathcal{K}'}^2\right] \geq \left[MV_{\mathcal{J}, \mathcal{K}}^2\right],$ with the concavity strict iff the inequality is.

If $k'$ is only weakly regular, then the calculations for the attribute case show that $\mathbb{E}\left[MV_{\mathcal{J},\mathcal{K}'}^2\right] \geq \mathbb{E}\left[MV_{\mathcal{J}, \mathcal{K}}^2\right],$ meaning that $\Delta \overline{W}$ must be at least weakly concave, in which case $\Delta \overline{W}(\delta) \leq 0$ for all $\delta > 0.$  In this case we set $\delta_* = 0$ and $\beta_* = \delta_*/(1 + \delta_*) = 0.$

Going forward, suppose that $k'$ is strictly regular.  Defining $\delta_* \equiv \inf\{\delta > 0 \ : \ \Delta \overline{W}(\delta) \leq 0\}$ yields a threshold in $(0, \infty]$ with the property that $\Delta \overline{W}(\delta) < 0$ for all $\delta > \delta_*$ and $\Delta \overline{W}(\delta) > 0$ for all $\delta \in (0, \delta_*).$  This threshold is finite iff $\Delta \overline{W}$ is strictly concave function.  Calculations very similar to the attribute case imply that
\begin{align*}
& \mathbb{E}[ MV_{\mathcal{J}, \mathcal{K}'}^2 \mid a_{\mathcal{J}}, c_{\mathcal{K}}] - MV_{\mathcal{J},\mathcal{K}}^2 \\
= \ & \mathbb{E}[MV_{\mathcal{J}, \mathcal{K}'} \mid a_{\mathcal{J}}, c_{\mathcal{K}}]^2 - MV_{\mathcal{J},\mathcal{K}}^2 + \text{Var}\left(MV_{\mathcal{J}, \mathcal{K}'} \mid a_{\mathcal{J}}, c_{\mathcal{K}}\right).
\end{align*}
Strict regularity implies that the difference of the first two terms on the rhs is non-negative for every realization of $(a_{\mathcal{J}}, c_{\mathcal{K}}),$  and positive with positive probability. Further, the conditional variance of $MV_{\mathcal{J}, \mathcal{K}'}$ must be non-negative.  Therefore
\[\mathbb{E}[ MV_{\mathcal{J}, \mathcal{K}'}^2 \mid a_{\mathcal{J}}, c_{\mathcal{K}}] \geq MV_{\mathcal{J},\mathcal{K}}^2,\]
with the inequality strict with positive probability.  Taking the unconditional expectation of both sides yields the desired strict concavity of $\Delta \overline{W},$ implying $\delta_* < \infty.$  Letting $\beta_* \equiv \delta_*/(1 + \delta_*)$ yields a reputation weight with the desired properties.

\subsection{Proof of Proposition \ref{prop:MVDiffCorr}}
We prove the first part of the result, with the second following from nearly identical arguments.  To streamline notation, throughout this proof we will drop superscripts on $Y^0$ and $\bar{Y}^0.$

Under Assumption \ref{ass:invertibilityCorrelated}, conditioning on the value of $X$ is equivalent to conditioning on the value of $\bar{Y}.$  The desired result can therefore be established using a technique very similar to the proof of Theorem \ref{thm:genBaseA}(a) subsequent to Lemma \ref{lemma:affiliationA2}, with $\bar{Y}$ playing the role of $\theta_{j'}$.  That argument requires two conditions: A) $(\bar{Y}, Y)$ are statistically affiliated, and B) $(\bar{Y}, \theta)$ are statistically affiliated conditional on $Y.$  We maintain condition A by hypothesis, so it remains only to establish condition B.

Let $\rho_{Y \mid \bar{Y}, \theta}(y \mid y', t)$ be the conditional density of $Y$ given $(\bar{Y}, \theta)$; $\rho_{\bar{Y}, \theta}$ be the joint density of $(\bar{Y}, \theta);$ and $\rho_Y$ be the marginal density of $Y.$  Then by Bayes' rule,
\[\rho_{\bar{Y}, \theta \mid Y}(y', t \mid y) = \frac{\rho_{Y \mid \bar{Y}, \theta}(y \mid y', t) \rho_{\bar{Y}, \theta}(y', t)}{\rho_Y(y)}.\]
Because $(\bar{Y}, \theta)$ are affiliated, the density $\rho_{\bar{Y}, \theta}$ is log-supermodular.  To establish condition B, it is therefore sufficient to show that $\rho_{Y \mid \bar{Y}, \theta}(y \mid y', t)$ is log-supermodular in $(y', t),$ holding $y$ fixed.

Let $\rho_{\eps \mid \bar{Y}, \theta}(e \mid y', t)$ and $\rho_{\eps \mid \bar{Y}}(e \mid y')$ be the conditional densities of $\eps$ given $(\bar{Y}, \theta)$ and $\bar{Y},$ respectively.  Given the conditional independence of $\theta$ and $\eps$ given $X$ and the invertibility of $\mathbb{E}(\bar{Y} \mid X),$ it follows that $\theta$ and $\eps$ are independent conditional on $\bar{Y}.$  Therefore
\[\rho_{\eps \mid \bar{Y}, \theta}(e \mid y', t) = \rho_{\eps \mid \bar{Y}}(e \mid y')\] for all $t,$ allowing us to write
\[\rho_{Y \mid \bar{Y}, \theta}(y \mid y', t) = \rho_{\eps \mid \bar{Y}, \theta}(y - t \mid y', t) = \rho_{\eps \mid \bar{Y}}(y - t \mid y').\]
Let $\rho_{\eps, \bar{Y}}(e, y')$ be the joint density of $(\eps, \bar{Y})$ and $\rho_{\bar{Y}}(y')$ be the marginal density of $\bar{Y}.$  Then using Bayes' rule, we have
\[\rho_{Y \mid \bar{Y}, \theta}(y \mid y', t) = \frac{\rho_{\eps, \bar{Y}}(y - t, y')}{\rho_{\bar{Y}}(y')}.\]
Condition B therefore follows if $\rho_{\eps, \bar{Y}}$ is log-submodular.  Equivalently, $\rho_{-\eps, \bar{Y}}$ must be log-supermodular, where $\rho_{-\eps,\bar{Y}}$ is the joint density of $(-\eps, \bar{Y}).$  This condition is equivalent to affiliation of $(-\eps, \bar{Y}),$ as hypothesized. 

\subsection{Proof of Lemma \ref{lemma:regularityGaussian}}
In a multivariate Gaussian environment the conditional mean of output is \[\bar{Y}^0 = \mu + (b + d)X,\] 
satisfying Assumption \ref{ass:invertibilityCorrelated} whenever $b + d \neq 0$.  Since $(X, Z, W)$ are jointly Gaussian and $\bar{Y}^0$ and $Y^0$ are each linear combinations of $(X, Z, W),$ the pair $(\bar{Y}^0, Y^0)$ are also jointly Gaussian.  Additionally,
\[Y^0 = \bar{Y}^0 + Z + W,\]
and since $\bar{Y}^0$ is independent of $Z + W$ it follows that $Y^0$ and $\bar{Y}^0$ are positively correlated.  Therefore $(\bar{Y}^0, Y^0)$ are affiliated, satisfying Assumption \ref{ass:regularityCorrelated}. Moreover, conditional expectations in multivariate Gaussian environments are linear in the conditioning variable, ensuring that Assumption \ref{ass:expectationDiffCorrelated} is satisfied.  Finally, mutual independence of $X$, $Z,$ and $W$ implies that $(\theta, \eps)$ are independent conditional on $X,$ satisfying Assumption \ref{ass:conditionalIndependence}. 

\subsection{Proof of Proposition \ref{prop:correlatedGaussian}}

Prior to the measurement, $(\theta, Y^0)$ have joint distribution
\[\left(\begin{array}{c}
\theta \\
Y^0
\end{array}\right) \sim \mathcal{N}\left(\left(\begin{array}{c}\mu\\ \mu\end{array}\right), 
\left(\begin{array}{cc}
\sigma_\theta^2 & \sigma_{\theta, Y} \\
\sigma_{\theta, Y} & \sigma_Y^2 \end{array}\right)\right)\]
where
\begin{align*}
    \sigma_\theta^2 = b^2 \sigma_x^2 + \sigma_z^2, \quad 
    \sigma_{\theta, Y} = b(b+d) \sigma_x^2 + \sigma_z^2, \quad
    \sigma_Y^2 = (b+d)^2\sigma_x^2 + \sigma_z^2 + \sigma_w^2. 
\end{align*}
The market's type forecast given equilibrium effort $e^*$ is therefore
\begin{align*}
   \mathbb{E}^{e^*}(\theta \mid Y) & =\mathbb{E}(\theta \mid Y^0 = Y - e^*) = \mu + \frac{\sigma_{\theta, Y}}{\sigma_Y^2}\cdot (Y - \mu - e^*),
\end{align*}
implying that the baseline marginal value of effort is
\begin{align*}
MV = \frac{\sigma_{\theta, Y}}{\sigma_Y^2} = \frac{b(b+d) \sigma_x^2 + \sigma_z^2}{(b+d)^2\sigma_x^2 + \sigma_z^2 + \sigma_w^2}.
\end{align*}
Meanwhile, after measuring $X$, the conditional joint distribution of $(\theta, Y^0)$ is 
\[\left(\begin{array}{c}
\theta \\ Y^0
\end{array}\right) \mid X \sim \mathcal{N}\left(\left(\begin{array}{cc}
\mu + b X \\
\mu + (b+d)X\end{array} \right), \left(\begin{array}{cc} \sigma_z^2 & \sigma_z^2 \\
\sigma_z^2 & \sigma_z^2 + \sigma_w^2 \end{array}\right)\right).\]
Given equilibrium effort $e^{**},$ the market's posterior type forecast after measuring $X$ is
\begin{align*}
\mathbb{E}^{e^{**}}(\theta \mid X, Y) = \mathbb{E}(\theta \mid X, Y^0 = Y - e^{**}) = \mu + bX + \frac{\sigma_z^2}{\sigma_z^2 + \sigma_w^2}(Y - e^{**} - \mu - (b + d)X).
\end{align*}
The agent's marginal value of effort under the expanded dataset is therefore
\[MV_+ = \frac{\sigma_z^2}{\sigma_z^2 + \sigma_w^2}.\]
Comparing the expressions for $MV$ and $MV_+$ just derived yields the identity in the proposition statement.

\phantomsection


\begin{thebibliography}{23}
\newcommand{\enquote}[1]{``#1''}
\expandafter\ifx\csname natexlab\endcsname\relax\def\natexlab#1{#1}\fi

\bibitem[\protect\citeauthoryear{Ball}{Ball}{2022}]{Ball}
\textsc{Ball, I.} (2022): \enquote{Scoring Strategic Agents,} Working Paper.

\bibitem[\protect\citeauthoryear{Bergemann, Bonatti, and Gan}{Bergemann
  et~al.}{2022}]{BergemannBonattiGan}
\textsc{Bergemann, D., A.~Bonatti, and T.~Gan} (2022): \enquote{The Economics
  of Social Data,} \emph{RAND Journal of Economics}, 53, 263--296.

\bibitem[\protect\citeauthoryear{Bonatti and Cisternas}{Bonatti and
  Cisternas}{2020}]{BonattiCisternas}
\textsc{Bonatti, A. and G.~Cisternas} (2020): \enquote{Consumer Scores and
  Price Discrimination,} \emph{The Review of Economic Studies}, 87, 750--791.

\bibitem[\protect\citeauthoryear{Braverman and Chassang}{Braverman and
  Chassang}{2022}]{Chassang}
\textsc{Braverman, M. and S.~Chassang} (2022): \enquote{Data-driven incentive
  alignment in capitation schemes,} \emph{Journal of Public Economics}, 207,
  104584.

\bibitem[\protect\citeauthoryear{Brunnermeier, Lamba, and
  Segura-Rodriguez}{Brunnermeier et~al.}{2021}]{LambaSegura}
\textsc{Brunnermeier, M., R.~Lamba, and C.~Segura-Rodriguez} (2021):
  \enquote{Inverse Selection,} Working Paper.

\bibitem[\protect\citeauthoryear{Dewatripont, Jewitt, and Tirole}{Dewatripont
  et~al.}{1999}]{DJT}
\textsc{Dewatripont, M., I.~Jewitt, and J.~Tirole} (1999): \enquote{The
  Economics of Career Concerns, Part I: Comparing Information Structures,}
  \emph{The Review of Economic Studies}, 66, 183–198.

\bibitem[\protect\citeauthoryear{Eliaz and Spiegler}{Eliaz and
  Spiegler}{2019}]{EliazSpiegler1}
\textsc{Eliaz, K. and R.~Spiegler} (2019): \enquote{The Model Selection Curse,}
  \emph{American Economic Review: Insights}, 1, 127--140.

\bibitem[\protect\citeauthoryear{Eliaz and Spiegler}{Eliaz and
  Spiegler}{2022}]{EliazSpiegler2}
---\hspace{-.1pt}---\hspace{-.1pt}--- (2022): \enquote{On Incentive-Compatible
  Estimators,} \emph{Games and Economic Behavior}, 122, 204--220.

\bibitem[\protect\citeauthoryear{Elliott, Galeotti, and Koh}{Elliott
  et~al.}{2021}]{ElliottGaleotti}
\textsc{Elliott, M., A.~Galeotti, and A.~Koh} (2021): \enquote{Market
  segmentation through information,} Working Paper.

\bibitem[\protect\citeauthoryear{Frankel and Kartik}{Frankel and
  Kartik}{2022}]{FrankelKartik}
\textsc{Frankel, A. and N.~Kartik} (2022): \enquote{Improving Information from
  Manipulable Data,} \emph{Journal of the European Economic Association}, 20,
  79--115.

\bibitem[\protect\citeauthoryear{Gomes and Pavan}{Gomes and
  Pavan}{2022}]{GomesPavan}
\textsc{Gomes, R. and A.~Pavan} (2022): \enquote{Price Customization and
  Targeting in Matching Markets,} \emph{RAND Journal of Economics},
  Forthcoming.

\bibitem[\protect\citeauthoryear{Haghtalab, Immorlica, Lucier, and
  Wang}{Haghtalab et~al.}{2020}]{Haghtalabetal}
\textsc{Haghtalab, N., N.~Immorlica, B.~Lucier, and J.~Wang} (2020):
  \enquote{Maximizing Welfare with Incentive-Aware Evaluation Mechanisms,} in
  \emph{Proceedings of the Twenty-Ninth International Joint Conference on
  Artificial Intelligence}, 160--166.

\bibitem[\protect\citeauthoryear{Hidir and Vellodi}{Hidir and
  Vellodi}{2021}]{HidirVellodi}
\textsc{Hidir, S. and N.~Vellodi} (2021): \enquote{Privacy, Personalization and
  Price Discrimination,} \emph{Journal of the European Economic Association},
  19, 1342–1363.

\bibitem[\protect\citeauthoryear{Holmstr\"{o}m}{Holmstr\"{o}m}{1999}]{Holmstrom}
\textsc{Holmstr\"{o}m, B.} (1999): \enquote{Managerial Incentive Problems: A
  Dynamic Perspective,} \emph{The Review of Economic Studies}, 66, 169--182.

\bibitem[\protect\citeauthoryear{Hu, Immorlica, and Vaughan}{Hu
  et~al.}{2019}]{Immorlica}
\textsc{Hu, L., N.~Immorlica, and J.~W. Vaughan} (2019): \enquote{The Disparate
  Effects of Strategic Manipulation,} in \emph{Proceedings of the Conference on
  Fairness, Accountability, and Transparency}, 259--268.

\bibitem[\protect\citeauthoryear{Ichihashi}{Ichihashi}{2019}]{Ichihashi}
\textsc{Ichihashi, S.} (2019): \enquote{Online Privacy and Information
  Disclosure by Consumers,} \emph{American Economic Review}, 110, 569--595.

\bibitem[\protect\citeauthoryear{Kearns and Roth}{Kearns and
  Roth}{2019}]{EthicalAlg}
\textsc{Kearns, M. and A.~Roth} (2019): \emph{The Ethical Algorithm: The
  Science of Socially Aware Algorithm Design}, Oxford University Press.

\bibitem[\protect\citeauthoryear{Meyer and Vickers}{Meyer and
  Vickers}{1997}]{MeyerVickers}
\textsc{Meyer, M.~A. and J.~Vickers} (1997): \enquote{Performance Comparisons
  and Dynamic Incentives,} \emph{Journal of Political Economy}, 105, 547--581.

\bibitem[\protect\citeauthoryear{Milgrom}{Milgrom}{1981}]{MilgromMLRP}
\textsc{Milgrom, P.} (1981): \enquote{Good News and Bad News: Representation
  Theorems and Applications,} \emph{The Bell Journal of Economics}, 12,
  380--391.

\bibitem[\protect\citeauthoryear{Milgrom and Roberts}{Milgrom and
  Roberts}{1988}]{influenceActivities}
\textsc{Milgrom, P. and J.~Roberts} (1988): \enquote{An Economic Approach to
  Influence Activities in Organizations,} \emph{American Journal of Sociology},
  94, S154--S179.

\bibitem[\protect\citeauthoryear{Rodina}{Rodina}{2018}]{rodina2018}
\textsc{Rodina, D.} (2018): \enquote{Information Design and Career Concerns,}
  Working Paper.

\bibitem[\protect\citeauthoryear{Tirole}{Tirole}{2021}]{Tirole}
\textsc{Tirole, J.} (2021): \enquote{Digital Dystopia,} \emph{American Economic
  Review}, 111, 2007--2048.

\bibitem[\protect\citeauthoryear{Yang}{Yang}{2022}]{YangJMP}
\textsc{Yang, K.~H.} (2022): \enquote{Selling Consumer Data for Profit: Optimal
  Market-Segmentation Design and its Consequences,} \emph{American Economic
  Review}, 112, 1364--1393.

\end{thebibliography}
\end{document}